\documentclass[]{mn2e}
\usepackage{multirow}
\usepackage{rotating}
\usepackage{colortbl}
\usepackage[fleqn]{amsmath}
\usepackage{amssymb}
\usepackage{amsfonts}
\usepackage{verbatim}
\usepackage{scalefnt}

\newcommand{\aj}{AJ}% Astronomical Journal
\newcommand{\araa}{ARA\&A}% Annual Review of Astron and Astrophys
\newcommand{\apj}{ApJ}% Astrophysical Journal
\newcommand{\apjl}{ApJ}% Astrophysical Journal, Letters
\newcommand{\apjs}{ApJS}% Astrophysical Journal, Supplement
\newcommand{\aap}{A\&A}% Astronomy and Astrophysics
\newcommand{\aapr}{A\&A~Rev.}% Astronomy and Astrophysics Reviews
\newcommand{\mnras}{MNRAS}% Monthly Notices of the RAS
\newcommand{\na}{New A}% New Astronomy
\newcommand{\pasj}{PASJ}% Publications of the ASJ
\newcommand{\aplett}{Astrophys.~Lett.}% Astrophysics Letters and Communications
\newcommand{\procspie}{Proc.~SPIE}% Proceedings of the SPIE

\newcommand{\comments}[1]{} %usage: \comments{}
\title[Polytropic approximation and scaling relations]{The polytropic approximation and X-ray scaling relations: constraints on gas and dark matter profiles for galaxy groups and clusters}

\author[P.R. Capelo, P.S. Coppi  and P. Natarajan]{Pedro R. Capelo,$^{1}$\thanks{E-mail: pedro.capelo@yale.edu} Paolo S. Coppi$^{1,2}$ and Priyamvada Natarajan$^{1,2}$\\
$^1$Department of Astronomy, Yale University, PO Box 208101, New Haven, CT 06520-8101, USA\\
$^2$Department of Physics, Yale University, PO Box 208120, New Haven, CT 06520-8120, USA}

\begin{document}

\maketitle

\begin{abstract}
The X-ray properties of groups and clusters of galaxies obey scaling relations that provide insight into the physics of their formation and evolution. In this paper, we constrain gas and dark matter parameters of these systems, by comparing the observed relations to theoretical expectations, obtained assuming that the gas is in hydrostatic equilibrium with the dark matter and follows a polytropic relation. In this exercise, we vary four parameters: the gas polytropic index $\Gamma$, its temperature at large radii, the dark matter logarithmic slope at large radii $\zeta$ and its concentration. When comparing the model to the observed mass-temperature relation of local high-mass systems, we find our results to be independent of both the gas temperature at large radii and of the dark matter concentration. We thus obtain constraints on $\Gamma$, by fixing the dark matter profile, and on $\zeta$, by fixing the gas profile. For a Navarro-Frenk-White dark matter profile, we find that $\Gamma$ must lie between 6/5 and 13/10. This value is consistent with numerical simulations and observations of individual clusters. Taking $6/5 \lesssim \Gamma \lesssim 13/10$ allows the dark matter profile to be slightly steeper than the Navarro-Frenk-White profile at large radii. Upon including local low-mass systems, we obtain constraints on the mass-dependence of $\Gamma$ and on the value of the gas temperature at large radii. Interestingly, by fixing $\Gamma=6/5$ and $\zeta=-3$, we reproduce the observed steepening/breaking of the mass-temperature relation at low masses if the temperature of the intercluster medium is between $10^6$ K and $10^7$ K, consistent with numerical simulations and observations of the warm-hot intergalactic medium. When extrapolated to high redshift, the model with a constant $\Gamma$ reproduces the expected self-similar behaviour. Given our formulation, we can also naturally account for the observed, non-self-similar relations provided by some high-redshift clusters, as they simply provide constraints on the evolution of $\Gamma$. In addition, comparing our model to the observed luminosity-temperature relation, we are able to discriminate between different mass-concentration relations and find that a weak dependence of concentration on mass is currently preferred by data. In summary, this simple theoretical model can still account for much of the complexity of recent, improved X-ray scaling relations, provided that we allow for a mild dependence of the polytropic index on mass or for a gas temperature at large radii consistent with intercluster values.

\end{abstract}


%SECTION 1
%SECTION 1
%SECTION 1
%SECTION 1
%SECTION 1

\section{Introduction}\label{sec:Introduction}

Clusters of galaxies are important probes of cosmology, as they complement other observations, including Type Ia supernov$\ae$ (e.g. Riess et al. 1998; Perlmutter et al. 1999) and the cosmic microwave background (e.g. Komatsu et al. 2009), in building the currently favoured ``concordance cosmology'' model. As the largest gravitationally bound systems in the Universe, galaxy clusters trace the growth of structure, which in turn strongly depends on the cosmological parameters (for recent reviews on clusters in a cosmological context, see e.g. Voit 2005; Allen, Evrard \& Mantz 2011). However, galaxy clusters can provide useful cosmological information only if their mass, composed predominantly of dark matter (DM), can be measured accurately enough. Given the elusive nature of DM, this can be achieved only through indirect methods, e.g. by measuring orbital velocities of galaxies (e.g. Zwicky 1933, 1937), by gravitational lensing techniques (e.g. Zwicky 1937), or through studying the intracluster medium (see reviews by e.g. Sarazin 1986; B{\"o}hringer \& Werner 2010).

The intracluster medium is the hot, X-ray emitting gas that to first order is in thermal equilibrium with the cluster gravitational potential. One of the most commonly used models for galaxy clusters utilises a King (1962) profile to describe the total gravitational potential with the isothermal gas in thermal equilibrium. Solving the hydrostatic equilibrium (HE) equation, the resulting profile, referred to as the $\beta$-model (Cavaliere \& Fusco-Femiano 1976, 1978), is widely utilised to describe radial profiles of clusters. However, detailed X-ray observations of individual clusters have shown clear radial gradients in their temperature profiles, hence calling into question the assumption of isothermality.

One of the simplest extensions of the isothermal model is to assume a polytropic relation, wherein gas density and pressure (or temperature) are related by a simple power law. This description naturally provides radial temperature gradients. A polytropic relation is indeed expected in ideal adiabatic processes, with the polytropic index in those cases being equal to the ratio of specific heats at constant pressure and constant volume. This value of the exponent, however, does not appear to be consistent with observations (e.g. Bode, Ostriker \& Vikhlinin 2009 and references therein), suggesting that the polytropic relation is not an equation of state. Although the origin of the observed polytropic relation is still a puzzle (but see Bertschinger 1985 for a possible explanation), HE-polytropic solutions have been considered since the early 1970s (e.g. Lea 1975). Hybrid descriptions have also been considered, where the polytropic relation and the $\beta$-model description for the gas density are employed simultaneously (Markevitch et al. 1998; Ettori 2000; Sanderson et al. 2003; Ascasibar et al. 2003).

Numerical prescriptions of DM have advanced to the point where we are fairly certain about the profile of the DM potential (e.g. Navarro, Frenk \& White 1996: NFW). Suto, Sasaki \& Makino (1998) first derived the HE-polytropic solution in the case of a gravitational potential described by an NFW profile. The concept is simple: once the DM has been completely described (in the case of NFW, by only two parameters, e.g. virial mass and concentration), the gas is analytically described with just three free parameters: the polytropic index and the values of two gas quantities, chosen amongst e.g. density, pressure and temperature, at some given radius (usually zero or infinity). Extracting these three gas parameters and checking results with observations is not entirely straightforward. Radial profiles of individual clusters (e.g. Pratt et al. 2007; Arnaud et al. 2010) can provide some hints, but resolution limits and scatter prevent us from obtaining accurate constraints on the three unknown scalars introduced above. Moreover, any uncertainty in the DM profile makes the constraining more difficult. More global scaling relations between cluster mass, temperature and luminosity, on the other hand, can be very helpful. If one has a model for the temperature profile, then one can also calculate the average temperature and obtain a theoretical mass-temperature ($M$--$T$) relation that can be easily compared to observations (e.g. Vikhlinin et al. 2006; Vikhlinin et al. 2009: V09; Sun et al. 2009: S09). This is what we largely pursue in this paper. Since these average quantities are usually computed after excising the central region of the cluster, any uncertainty in the properties of the inner parts of the DM halo becomes unimportant. On the other hand, any change in the outer regions of the cluster can produce significant effects on the average quantities. For this reason, we do not limit ourselves to the NFW profile, but use its generalised version by Bulbul et al. (2010: B10), in which the DM logarithmic slope at large radii is not fixed. For completeness, given the uncertainty in the mass-concentration relation, we also let the concentration vary, thus having two DM free parameters.

The choice of the three primary gas quantities and the method for constraining them vary between authors. Komatsu \& Seljak (2001: KS01), in their description of a universal gas profile, assume that the gas density profile is equal to that of DM in the outer regions of the DM halo (assumed to be described by an NFW profile), obtaining a constraint on both the polytropic index and the central gas temperature. Their assumption is reasonable and very powerful, as it gives constraints on the polytropic index itself, which has not been done in other studies. However, it predicts an $M$--$T$ relation which is not in good agreement with recent X-ray observations. Also, the gas profile is not completely described, as it needs a second constraint to fix the central density. Komatsu \& Seljak (2002: KS02) provide such a constraint by requiring that the gas density in the outer regions of the DM halo is equal to the DM density multiplied by the cosmic baryon fraction.

Ascasibar et al. (2003), investigating the differences between polytropic relations and $\beta$-models, fix instead the polytropic index as an output from their numerical simulations and obtain constraints on the central gas temperature and density, by imposing the gas density to be zero at infinity and the baryon fraction to never exceed the cosmic value. Ascasibar et al. (2006: A06), studying the origin of self-similar scaling relations, use a similar method, with the slight difference that the baryon fraction at three times the DM scale radius must be equal to the cosmic value.

More recently, a series of papers (Ostriker, Bode \& Babul 2005; Bode et al. 2007; Bode et al. 2009) have explored in detail models with polytropic gas in HE with a DM potential, adding additional physics such as star formation and feedback, in the form of simple recipes. Similar to the approach of Ascasibar et al. (2003) and A06, Ostriker et al. (2005) take the polytropic index as a given (from numerical simulations and observations) and additionally constrain the central gas pressure and density by imposing conservation of energy and by matching the external surface pressure to the momentum flux from the infalling gas at the virial radius. The second constraint arises from a sharp truncation in the DM profile at the virial radius, adopted by Ostriker et al. (2005), but not contemplated in KS01, KS02, Ascasibar et al. (2003), A06, nor in this paper. By adding terms to the energy conservation equation, they are able to discriminate between simple feedback recipes when matching the resulting models to observations.

In this paper, we assume that an ideal polytropic gas is in HE with the DM in groups and clusters of galaxies and explore the consequences of the polytropic solution, considering different ways to specify the three primary gas quantities and the two free DM parameters. This is done assuming no prior knowledge of either the polytropic index or the boundary conditions, and only partial knowledge of the DM profile. Thus, our approach differs from previous work in the following ways. (i) We do not assume the DM to be NFW-like at large radii; instead, we adopt a generalised version of the NFW profile (B10), and are able to discriminate between different values of the logarithmic slope of the DM at large radii. (ii) We do not fix the value of the polytropic index from simulations (Ascasibar et al. 2003; A06; Ostriker et al. 2005; Bode et al. 2007, 2009) nor constrain it with some assumption on the gas profile at large radii (KS01; KS02); instead, we leave the polytropic index as a free parameter and, by comparing our model to observations, constrain it as a function of mass and redshift. (iii) We do not force the gas radial profile to behave in any way, either by imposing its profile to be equal to that of DM at large radii (KS01; KS02) or by requiring it to vanish at infinity (Ascasibar et al. 2003; A06); instead, we simply require the temperature of the gas to never be negative. By leaving the gas temperature at large radii as a free parameter, we are able to constrain the value of the temperature of the intracluster medium (or of the intercluster medium, depending on the assumptions we make) by matching our theoretically derived $M$--$T$ relation to observations in the low-mass regime (i.e. groups). (iv) We do not fix the concentration from simulations (KS01; KS02; Ascasibar et al. 2003; A06; Ostriker et al. 2005; Bode et al. 2007, 2009), but instead investigate the effects of different mass-concentration relations; we discriminate between different recipes by matching our model to recently observed luminosity-temperature ($L\,$--$\,T$) relations (Maughan et al. 2011: M11). For this last part of the study, a second constraint was necessary, and we imposed the baryon fraction within the virial radius to be equal to the cosmic value.

We believe the flexibility of our model to be its main strength. More general assumptions than in previous studies help us obtain more constraints: on the polytropic index; on the DM logarithmic slope at large radii; on the gas temperature at large radii; on concentration. Also, we do extend the model to higher redshift and obtain constraints on parameter evolution.

Throughout the paper, we use the following cosmological parameters (WMAP5; Komatsu et al. 2009): $\Omega_{\rm 0}=0.258$, $\Omega_{\rm b}=0.0441$, $\Omega_{\rm \Lambda}=0.742$, $H_{\rm 0}=100\,h$ km s$^{-1}$ Mpc$^{-1}=70\,h_{\rm 70}$ km s$^{-1}$ Mpc$^{-1}$, $h=0.719$ (i.e. $h_{\rm 70}=1.027$). The values and uncertainties of other parameters (e.g. $\sigma_{\rm 8}$) are not relevant in our study.

The outline of the paper is as follows. In Section \ref{sec:Polytropic_gas_in_galaxy_groups_and_clusters}, we formulate the problem and study in detail the consequences of assuming hydrostatic equilibrium and a polytropic relation. In Section \ref{sec:Scaling_relations_and_comparison_to_observations}, we derive galaxy groups and clusters scaling relations and compare them to recent observations to obtain constraints on gas and DM parameters. We conclude in Section \ref{sec:Conclusions}.

%SECTION 2
%SECTION 2
%SECTION 2
%SECTION 2
%SECTION 2

\section{Polytropic gas in galaxy groups and clusters}\label{sec:Polytropic_gas_in_galaxy_groups_and_clusters}

%SECTION 2.1
%SECTION 2.1
%SECTION 2.1
%SECTION 2.1
%SECTION 2.1

\subsection{Formulation}\label{sec:Formulation}

In this section, we model groups and clusters of galaxies as isolated, spherically symmetric, composite systems of stars, DM and ideal gas. We describe the DM with a generalised NFW density profile (B10),

\begin{equation}\label{eq:DM_rho}
\rho_{\rm DM}(r) = \frac{\delta_{\rm c} \rho_{\rm c}}{(r/r_{\rm s})(1+r/r_{\rm s})^{\beta}},
\end{equation}

\noindent where $1 < \beta \leq 3$, $r_{\rm s}$ is the scale radius, $\delta_{\rm c}$ is the characteristic (dimensionless) density, $\rho_{\rm c}(z)=3H_{\rm 0}^2E^2(z)/(8\pi G)$ is the critical density of the Universe at redshift $z$, $E(z) \equiv H(z)/H_{\rm 0} \equiv [\Omega_{\rm 0}(1+z)^3+(1-\Omega_{\rm 0}-\Omega_{\rm \Lambda})(1+z)^2+\Omega_{\rm \Lambda}]^{1/2}$ and $G$ is the gravitational constant\footnote{This profile and $\beta$ should not be confused with the $\beta$-model of Cavaliere \& Fusco-Femiano (1976, 1978).}. Note that the logarithmic slope at small radii is the same as for the standard NFW profile, which is a particular case of this profile when $\beta=2$. We further assume the gas to follow a polytropic relation, with the pressure\footnote{Unless otherwise stated, pressure is thermal gas pressure, density is gas mass per unit volume, and temperature is gas temperature.} $P$, density $\rho$ and temperature $T$ related as follows: $P(r)=P_{\rm 0}[\rho(r)/\rho_{\rm 0}]^{\Gamma}$ or, equivalently, $\rho(r)=\rho_{\rm 0}[T(r)/T_{\rm 0}]^{1/(\Gamma-1)}$, where $P_{\rm 0}$, $\rho_{\rm 0}$ and $T_{\rm 0}$ are the central values of pressure, density and temperature, respectively, and $d(\log P)/d(\log \rho) \equiv \Gamma > 1$ is the polytropic index\footnote{In other notations, the polytropic index $n$ is related to $\Gamma$ via $n=1/(\Gamma-1)$.}, assumed to be the same at all radii (i.e. the polytropic relation is \emph{complete}).

We assume the gas to be in global, thermally-supported HE with the total gravitational potential $\phi$, that is, to obey the HE equation, $\nabla P=-\rho \nabla \phi$, everywhere. We also neglect the effects of gas self-gravity, as it has been demonstrated (e.g. Suto et al. 1998) that the contribution of the gas to the total gravitational potential does not lead to significantly different results, and additionally ignore the stellar gravitational potential, as it has been shown (see Capelo, Natarajan \& Coppi 2010 for the low-mass case) that gas profiles are affected by stars only in the very inner regions, which are not important for the results of this work\footnote{To be precise, the stellar gravitational potential has a strong effect on the central gas quantities (e.g. Capelo et al. 2010), but not on the luminosity and average temperature, when calculated with an excision of the central region (see Section \ref{sec:Scaling_relations_and_comparison_to_observations}). For the same reason, we do not investigate DM profiles with different logarithmic slopes at small radii (e.g. Moore et al. 1998).}. The total gravitational potential is thus dominated by that of DM, i.e. $\phi=\phi_{\rm DM}$, which is given by

\begin{equation}\label{eq:DM_phi}
\phi_{\rm DM}(r) =
         \begin{cases}\displaystyle{\phi_{\rm 0} \frac{1}{\beta -2}\frac{(1+r/r_{\rm s})^{\beta -2}-1}{(r/r_{\rm s})(1+r/r_{\rm s})^{\beta -2}}} & \mbox{if $\beta \neq 2$,} \cr
                      \displaystyle{\phi_{\rm 0}\ln[1+(r/r_{\rm s})]/(r/r_{\rm s})} & \mbox{if $\beta = 2$,} \cr
         \end{cases}
\end{equation}

\noindent where $\phi_{\rm 0} \equiv \phi_{\rm DM}(0) = -4\pi G\delta_{\rm c}\rho_{\rm c}r_{\rm s}^2/(\beta-1)$ and we have used Equation \eqref{eq:DM_rho} to solve Poisson's equation, $\nabla^2 \phi_{\rm DM} = 4\pi G \rho_{\rm DM}$, after imposing the Dirichlet boundary condition $\phi_{\rm DM}(+\infty) = 0$.

The solution to the HE equation is then given by (Suto et al. 1998)

\begin{equation}\label{eq:polytropic_rho_NFW}
\rho(r) = \rho_{\rm 0} \left[1-\frac{\Gamma-1}{\Gamma}\Delta_{\rm gas}\left(1-\frac{\ln(1+r/r_{\rm s})}{r/r_{\rm s}}\right)\right]^{\frac{1}{\Gamma-1}}
\end{equation}

\noindent for $\beta = 2$ and by

\scalefont{0.94}
\begin{equation}\label{eq:polytropic_rho_B10}
\rho(r) = \rho_{\rm 0} \left[1-\frac{\Gamma-1}{\Gamma}\Delta_{\rm gas}\left(1-\frac{1}{\beta -2}\frac{\left(1+\frac{r}{r_{\rm s}}\right)^{\beta -2}-1}{\frac{r}{r_{\rm s}}\left(1+\frac{r}{r_{\rm s}}\right)^{\beta -2}}\right)\right]^{\frac{1}{\Gamma-1}}
\end{equation}
\normalsize

\noindent for $\beta \neq 2$, where the dimensionless gas parameter

\begin{equation}\label{eq:gas_parameter}
\Delta_{\rm gas} = -\phi_{\rm 0}\frac{\rho_{\rm 0}}{P_{\rm 0}} = \frac{4\pi G \delta_{\rm c} \rho_{\rm c} r_{\rm s}^2 \mu m_{\rm p}}{k_{\rm B}T_{\rm 0}(\beta-1)},
\end{equation}

\noindent $k_{\rm B}$ is the Boltzmann constant, $m_{\rm p}$ is the proton mass and $\mu = 0.62$ is the mean molecular weight of the gas (e.g. Parrish et al. 2011), assumed to be completely ionized and to have mean metallicity [Fe/H] $=-0.5$.

In both cases, we used the boundary condition $\rho(0)\equiv\rho_{\rm 0}$, where $\rho_{\rm 0}$ can in principle have any non-negative value (although we will see later in this section how to constrain it). B10 imposed a more stringent boundary condition, $T(+\infty) \equiv T_{\rm \infty} = 0$, and obtained a solution which is a particular case of Equation \eqref{eq:polytropic_rho_B10} when $(\Gamma -1)\Delta_{\rm gas} /\Gamma = 1$. Our solution is a more general case that allows for any non-negative value of $T_{\rm \infty}$, which can be then fixed depending on the physical situation (see Section \ref{sec:Dependence_on_the_gas_temperature_at_large_radii} for more details).

We now specify the problem further and consider a system with a given virial mass $M_{\rm vir}$ at a given redshift $z$, assuming that the virialisation was reached right before the time at which we observe it (recent-formation approximation). In this paper, unless otherwise stated, we define virial mass as the total (stars, DM and gas) mass within the virial radius, defined below. The DM density profile given in Equation \eqref{eq:DM_rho} can then be written, via the equality $\delta_{\rm c} = f_{\rm DM} \Delta_{\rm vir}c_{\rm vir}^3/[3f(c_{\rm vir})]$, as

\begin{equation}\label{eq:DM_rho_alternative}
\rho_{\rm DM}(r) = \frac{M_{\rm DM}(r_{\rm vir})}{4\pi f(c_{\rm vir})}\frac{r_{\rm s}^{\beta-2}}{r(r+r_{\rm s})^{\beta}},
\end{equation}

\noindent where the virial radius $r_{\rm vir}=[3M_{\rm vir}/(4\pi \Delta_{\rm vir}\rho_{\rm c})]^{1/3}$ and the concentration $c_{\rm vir}=r_{\rm vir}/r_{\rm s}$ are both assumed to be independent of $\beta$; $\Delta_{\rm vir}$ is the virial overdensity, described by $\Delta_{\rm vir}(z)\simeq18\pi^2+82[\Omega_{\rm m}(z)-1]-39[\Omega_{\rm m}(z)-1]^2$ (Bryan \& Norman 1998); the matter fraction $\Omega_{\rm m}(z)=\Omega_{\rm 0}(1+z)^3/E^2(z)$; $f_{\rm DM} = M_{\rm DM}(r_{\rm vir})/M_{\rm vir} = 1-f_{\rm b}$ is the virial (that is, within the virial radius) DM mass fraction; $f_{\rm b} = \Omega_{\rm b}/\Omega_{\rm 0}$ is the virial baryon mass fraction, set equal to the cosmic value, assumed constant with redshift and mass; the function $f(x)=\ln(1+x)-x/(1+x)$, when $\beta = 2$, or $f(x)=[1/(\beta -1)+[-x+1/(1-\beta)]/(1+x)^{\beta -1}]/(\beta -2)$, when $\beta \neq 2$. We note that several authors use different definitions for mass and radius. In general, one can define $r_{\rm \Delta_{\rm c}}$ such that $M_{\rm \Delta_{\rm c}}=4\pi r_{\rm \Delta_{\rm c}}^3 \Delta_{\rm c} \rho_{\rm c}/3$, where $\Delta_{\rm c}$ is the overdensity ratio. This ratio can be either a fixed number (e.g. 200, 500, etc.) or can be replaced by the just defined $\Delta_{\rm vir}$, so that $r_{\rm \Delta_{\rm c}}$ is the virial radius $r_{\rm vir}$ and $M_{\rm \Delta_{\rm c}}$ is the virial mass $M_{\rm vir}$. See Appendix \ref{sec:Overdensities} for the calculation of the relations between different overdensity masses and radii, for both the NFW case and for the more general B10 case. Moreover, some authors use equivalent but different notations for the virial overdensity: $M_{\rm vir}=4\pi r_{\rm vir}^3\Delta_{\rm vir}\Omega_{\rm m}(z)\rho_{\rm c}/3$ and $\Delta_{\rm vir}(z)\simeq[18\pi^2+82[\Omega_{\rm m}(z)-1]-39[\Omega_{\rm m}(z)-1]^2]/\Omega_{\rm m}(z)$.

The dependence of the concentration on redshift and virial mass can be parametrized as

\begin{equation}\label{eq:c_vir_M_vir_relation}
c_{\rm vir} = c_{\rm A}\left(\frac{f_{\rm DM}M_{\rm vir}}{M_{\rm c}}\right)^{c_{\rm B}}(1+z)^{c_{\rm C}},
\end{equation}

\noindent where the dimensionless parameters $c_{\rm A}, c_{\rm B}, c_{\rm C}$ and the scale mass $M_{\rm c}$ depend on the cosmological parameters. For our fiducial model and the rest of this section, we use results from recent DM-only simulations of relaxed haloes ($c_{\rm A}=9.23, c_{\rm B}=-0.09, c_{\rm C}=-0.69, M_{\rm c}=2\times10^{12}h^{-1}$M$_{\rm \odot}$; Duffy et al. 2008: D08)\footnote{We assume Equation \eqref{eq:c_vir_M_vir_relation} to be valid for any $1<\beta\leq 3$ and for any $10^{12}$M$_{\rm \odot} \leq M_{\rm vir} \leq 10^{16}$M$_{\rm \odot}$.}, which used WMAP5 cosmological parameters, but we will consider other published relations in Section \ref{sec:Scaling_relations_and_comparison_to_observations}.

Finally, one calculates the central density $\rho_{\rm 0}$ by imposing

\begin{equation}\label{eq:constraint_on_rho_gas}
\int_0^{r_{\rm vir}}\rho(r)4\pi r^2 dr=f_{\rm gas}M_{\rm vir}=(f_{\rm b}-f_{\rm star})M_{\rm vir},
\end{equation}

\noindent where $f_{\rm gas}$ and $f_{\rm star}$ are the virial gas and stellar mass fraction, respectively. For $f_{\rm star}$, we will consider two limiting cases:

\begin{equation}\label{eq:f_star}
f_{\rm star}=5.7\times10^{-2}\left(\frac{M_{\rm vir}}{5\times 10^{13}M_{\rm \odot}}\right)^{-0.26},
\end{equation}

\noindent following the work of Giodini et al. (2009)\footnote{We approximate their results for the full sample of COSMOS selected groups only, extrapolating to higher masses, where the stellar fraction becomes negligible anyway, and assuming that the stellar fraction within $r_{\rm 500}$ is equal to the stellar fraction within $r_{\rm vir}$. This last point is an approximation, since we know that the gas mass increases with radius, but it does so only slightly.}, or $f_{\rm star}=0$, depending on whether we want to quantify the effect (or lack thereof) on the mass budget of stars in low-mass systems. Since we are neglecting gas self-gravity, the choice of $\rho_{\rm 0}$ (or, equivalently, of $f_{\rm star}$) does not affect the results on the $M$--$T$ relation in Section \ref{sec:The_M_T_Relation}. However, it does affect the results on the $L\,$--$\,T$ relation in Section \ref{sec:The_L_T_Relation}. For the remainder of this section, we will assume $f_{\rm star}=0$ (i.e. $f_{\rm gas}=f_{\rm b}$).

All parameters but one have now been fixed. The central pressure or, alternatively, the central temperature can in principle have any value. In reality, there is another constraint that helps us fix $T_{\rm 0}$.

A general attribute of the polytropic solution of the HE equation is that it is physically meaningful only between zero and a maximum radius $r_{\rm max}$ (Lea 1975). Beyond $r_{\rm max}$, the temperature becomes negative, making the solution unphysical in that region\footnote{For some values of $\Gamma$ (e.g. $\Gamma=3/2$), the density is positive also for $r>r_{\rm max}$, but the temperature becomes negative regardless of the value of the polytropic index.}. The existence of a maximum radius can be explained in the same way one explains the finite truncation radius in classical polytropes with $\Gamma > 6/5$: a finite $r_{\rm max}$ means that the gravitational potential is strong enough to bind the gas in a finite region; an infinite $r_{\rm max}$ means that the pressure of the gas is high enough to make the gas either unbound or bound only in a limitless region. However, the analogy between the profiles in this work and those of classical polytropes cannot be taken too far: in classical polytropes, $r_{\rm max}=+\infty$ if $\Gamma \leq 6/5$, regardless of other gas quantities; on the other hand, the polytropic solution given by Equation \eqref{eq:polytropic_rho_NFW} or \eqref{eq:polytropic_rho_B10}, for a given polytropic index, can have a finite or infinite maximum radius, depending on other gas quantities, as it is explained below. The difference arises from the different nature of $\phi$: in this paper, the gravitational potential is fixed and external, due to DM; in the classical polytrope case, $\phi$ is entirely due to gas self-gravity.

By studying Equations \eqref{eq:polytropic_rho_NFW} and \eqref{eq:polytropic_rho_B10} further, we note that $r_{\rm max}$, once all the other parameters (virial mass, redshift, cosmological parameters, etc.) have been fixed, depends only on the gas temperature. The hotter the gas is, the larger $r_{\rm max}$ becomes. In particular, by imposing the weak boundary condition $T_{\rm \infty} \geq 0$, there exists a threshold minimum value of the central temperature,

\begin{equation}\label{eq:T_0_thr}
T_{\rm 0-thr}=\frac{4\pi G \delta_{\rm c} \rho_{\rm c} r_{\rm s}^2 \mu m_{\rm p}}{k_{\rm B}(\beta -1)} \frac{\Gamma-1}{\Gamma}=-\phi_{\rm 0}\frac{\Gamma-1}{\Gamma}\frac{\mu m_{\rm p}}{k_{\rm B}},
\end{equation}

\noindent above which the solution is physically meaningful at all radii, i.e. $r_{\rm max}=+\infty$. This central threshold temperature, which depends on the system's gravitational potential (i.e. on its mass) and on the steepness of the polytropic relation, is a characteristic quantity of the system: $T_{\rm 0-thr}$ is the minimum permitted central temperature, resulting from the requirement of HE and non-negative temperature at all radii. As expected, a deeper gravitational potential requires a higher threshold central temperature. There exists also a maximum permitted central temperature, constrained by the boundary condition at large radii, usually given by the temperature of the intracluster medium or of the intercluster medium, depending on the type of system (e.g. group vs. cluster; isolated vs. non-isolated) under consideration. There is therefore, for any given $\Gamma$, a relation between the mass of a system and a range of permitted central temperatures of its gas. This $M$--$T$ relation will be fully exploited in Section \ref{sec:Scaling_relations_and_comparison_to_observations} to constrain gas and DM parameters. If $T_{\rm 0}<T_{\rm 0-thr}$, $r_{\rm max}$ is finite and can even be much smaller than the virial radius, depending on the other parameters. If $T_{\rm 0}=T_{\rm 0-thr}$, $r_{\rm max}=+\infty$ and all gas quantities at infinity vanish (and we recover the solution given in B10). If $T_{\rm 0}>T_{\rm 0-thr}$, $r_{\rm max}=+\infty$ and all gas quantities at infinity are positive. Imposing the boundary condition $T(+\infty)=T_{\rm \infty}$, we have $T_{\rm 0}=T_{\rm 0-thr}+T_{\rm \infty}$.

\begin{figure}
\centering
\vspace{5pt}
\includegraphics[width=0.74\columnwidth,angle=90]{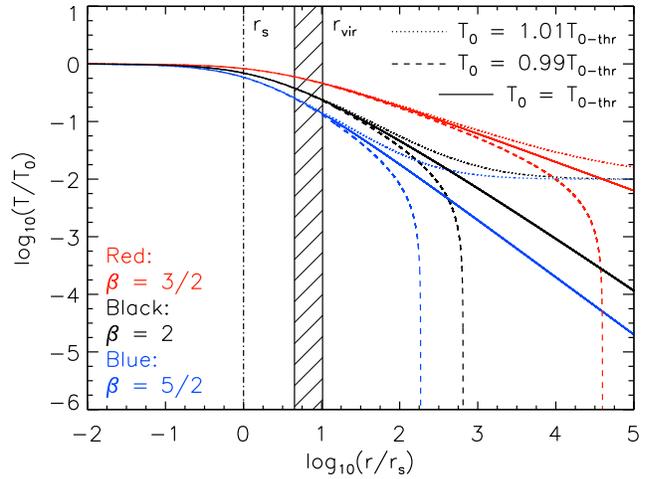}
\vspace{-8pt}
\caption{Dependence of the polytropic solution to the HE equation (Equations \ref{eq:polytropic_rho_NFW} and \ref{eq:polytropic_rho_B10}) on the gas central temperature, for three different values of $\beta$: 3/2 (\emph{upper, red bundle of curves}), 2 (NFW case; \emph{middle, black bundle of curves}) and 5/2 (\emph{lower, blue bundle of curves}). The \emph{black, vertical, dot-dashed line} denotes the DM scale radius $r_{\rm s}$, whereas the \emph{vertical band} denotes the virial radius, for a range of systems at redshift $z=0$ and with $10^{12}$M$_{\rm \odot} \leq M_{\rm vir} \leq 10^{16}$M$_{\rm \odot}$. The temperature profile depends strongly on the value of $\beta$. When $T_{\rm 0}=T_{\rm 0-thr}$ (\emph{solid curves}), the logarithmic slope at large radii $d(\log T)/d(\log r)$ is $\simeq -1$ for $\beta \geq 2$ and $\simeq -\beta+1$ for $\beta < 2$. Additionally, at very large radii there is a strong dependence on $T_{\rm 0}$: the \emph{dotted} (\emph{dashed}) \emph{curves} show the temperature profile when the central temperature is increased (decreased) by 1 per cent from $T_{\rm 0-thr}$.}
\label{fig:T_Radial_Profile}
\end{figure}

In Figure \ref{fig:T_Radial_Profile}, we show the effect of changing $T_{\rm 0}$ on the temperature radial profile, derived from Equations \eqref{eq:polytropic_rho_NFW} and \eqref{eq:polytropic_rho_B10} and from the polytropic relation, for a few cases of $\beta$. The solution with $T_{\rm 0}=T_{\rm 0-thr}$ is the only case for which the temperature becomes zero at infinity, and it does so with a logarithmic slope at large radii $d(\log T)/d(\log r) \simeq -1$ for $\beta \geq 2$ and $\simeq -\beta+1$ for $\beta < 2$. This strong dependence on $\beta$ will be useful in Section \ref{sec:Dependence_on_the_DM_logarithmic_slope_at_large_radii}, where we constrain the DM logarithmic slope at large radii $\zeta \equiv -\beta -1$. Notice also that, in this particular case, the shape of the profile is independent of the polytropic index. When $T_{\rm 0} \neq T_{\rm 0-thr}$, the temperature profile $T(r)/T_{\rm 0}$ depends on the polytropic index. However, when $\Gamma$ varies, so do $T_{\rm 0-thr}$ and its fractions (e.g. $0.99\,T_{\rm 0-thr}$ and $1.01\,T_{\rm 0-thr}$, shown in the figure), therefore Figure \ref{fig:T_Radial_Profile} looks exactly the same for any value of $\Gamma > 1$. If we slightly increase or decrease the central temperature, the effect on $T(r)/T_{\rm 0}$ is quite dramatic at very large radii, whereas it is almost not observed within the virial radius.

\begin{figure}
\centering
\vspace{5pt}
\includegraphics[width=0.7\columnwidth,angle=90]{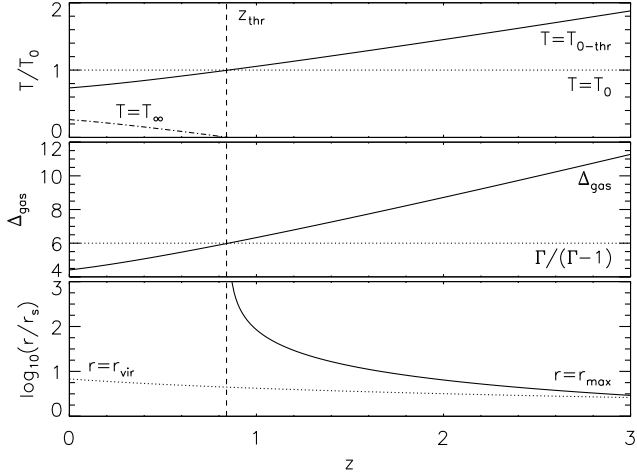}
\vspace{5pt}
\caption{Dependence on redshift of several parameters, for a cluster of virial mass $M_{\rm vir}=10^{14}$ M$_{\rm \odot}$, composed of NFW DM and gas with polytropic index $\Gamma=6/5$ and central temperature $k_{\rm B}T_{\rm 0}=2.5$ keV. {\bf Top panel}: we plot the temperature of the cluster $T_{\rm 0}$ (\emph{dotted line}), its central threshold temperature $T_{\rm 0-thr}$ (\emph{solid line}) and its temperature at infinity $T_{\rm \infty}$ (\emph{dot-dashed line}). Notice how $k_{\rm B}T_{\rm 0-thr}$ does not depend very strongly on redshift, varying from $\sim2$ keV to $\sim5$ keV in the $z=0\,$--$\,3$ range. {\bf Middle panel}: we plot the gas parameter $\Delta_{\rm gas}$ (\emph{solid line}) in relation to the ratio $\Gamma/(\Gamma-1)$ (\emph{dotted line}). {\bf Bottom panel}: we plot the virial radius $r_{\rm vir}$ (\emph{dotted line}) and the maximum radius $r_{\rm max}$ (\emph{solid line}) of the cluster. {\bf All panels}: the \emph{dashed, vertical line} denotes the threshold redshift $z_{\rm thr}$ at which $T_{\rm 0}=T_{\rm 0-thr}$, $T_{\rm \infty}=0$, $\Delta_{\rm gas}=\Gamma/(\Gamma-1)$ and the function $r_{\rm max}(z)$ has its vertical asymptote. When we impose $T_{\rm \infty}=0$, $z=z_{\rm thr}$ is the only redshift at which the cluster can exist in HE at all radii. If we instead allow for any $T_{\rm \infty} \geq 0$, then the cluster can exist in HE at all radii in the $z=0\,$--$\,z_{\rm thr}$ range.}
\label{fig:Dependence_on_z}
\end{figure}

\begin{figure}
\centering
\vspace{5pt}
\includegraphics[width=0.7\columnwidth,angle=90]{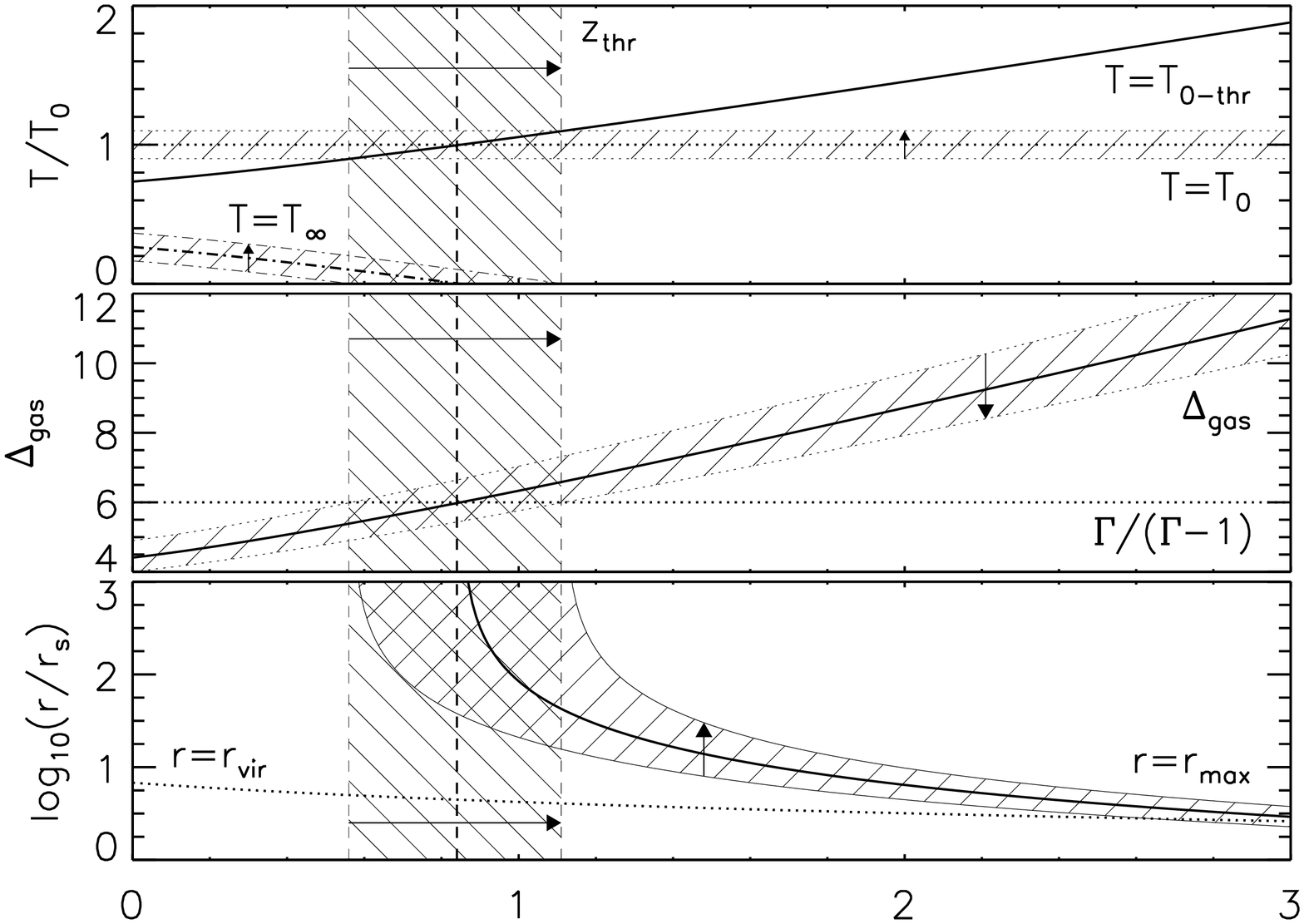}
\vspace{5pt}
\caption{Dependence of the parameters of the cluster described in Figure \ref{fig:Dependence_on_z} on its gas central temperature. The thick lines are the same lines shown in Figure \ref{fig:Dependence_on_z}. The arrows and the bands show what changes when we vary the central temperature from $0.9\,T_{\rm 0}$ to $1.1\,T_{\rm 0}$. When increasing the central temperature, $\Delta_{\rm gas}$ decreases, $T_{\rm \infty}$ and $r_{\rm max}$ increase and the threshold redshift $z_{\rm thr}$ increases. Notice the existence of a \emph{minimum} central temperature $T_{\rm 0-min}$, below which the cluster cannot exist in HE at all radii, at any redshift.}
\label{fig:Dependence_on_z_varying_T}
\end{figure}

\begin{figure}
\centering
\vspace{5pt}
\includegraphics[width=0.7\columnwidth,angle=90]{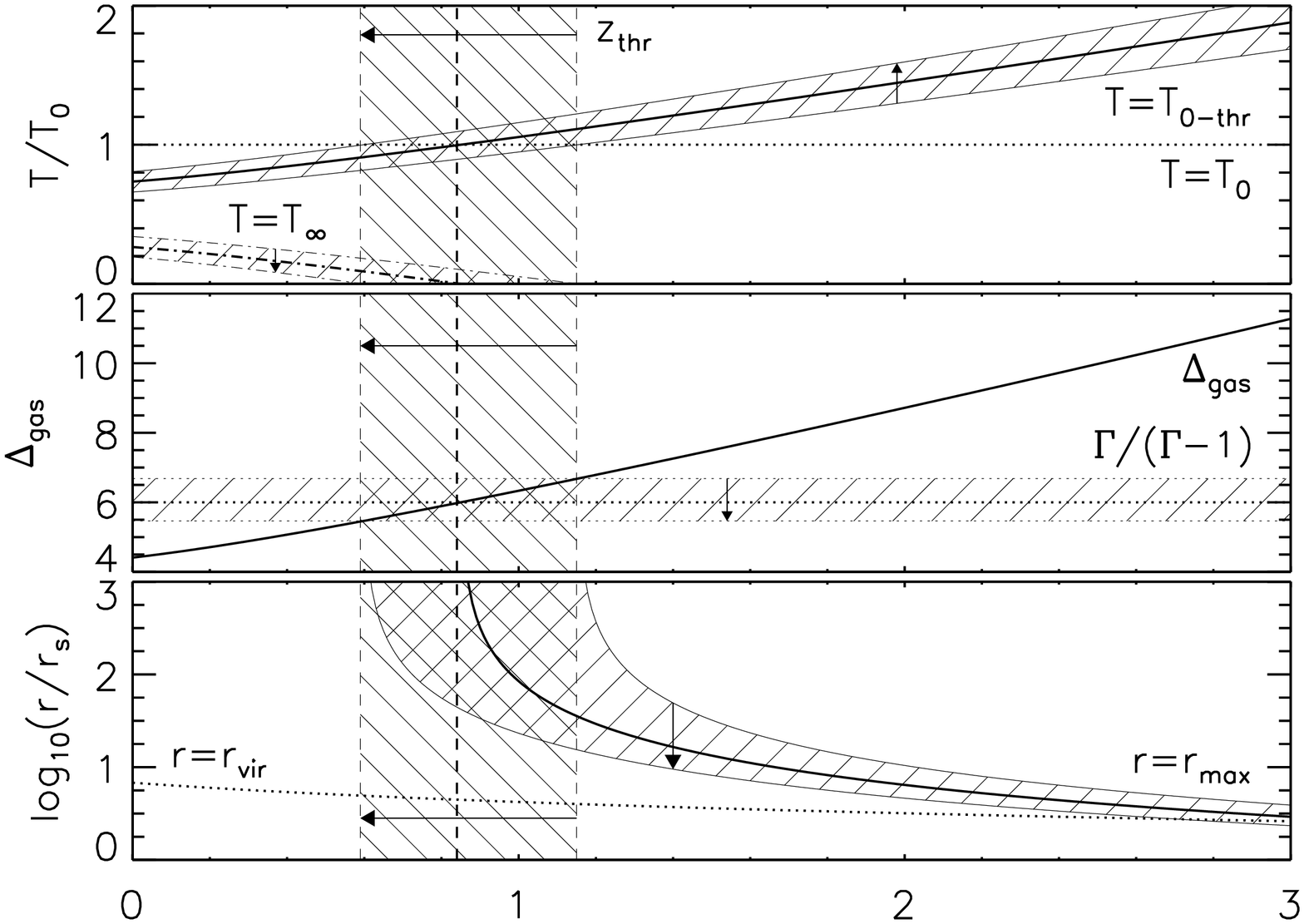}
\vspace{5pt}
\caption{Dependence of the parameters of the cluster described in Figure \ref{fig:Dependence_on_z} on its gas polytropic index. The thick lines are the same lines shown in Figure \ref{fig:Dependence_on_z}. The arrows and the bands show what changes when we vary the polytropic index from $0.98\,\Gamma$ to $1.02\,\Gamma$. When increasing the polytropic index, $T_{\rm 0-thr}$ increases, $T_{\rm \infty}$ and $r_{\rm max}$ decrease and the threshold redshift $z_{\rm thr}$ decreases. Notice the existence of a \emph{maximum} polytropic index $\Gamma_{\rm max}$, above which the cluster cannot exist in HE at all radii, at any redshift.}
\label{fig:Dependence_on_z_varying_Gamma}
\end{figure}

We can calculate the central temperature when one does not impose $r_{\rm max}=+\infty$, but imposes instead a finite $r_{\rm max}=\kappa \,r_{\rm vir}$, where $\kappa$ is some large integer. In this case, the new central temperature $T_{\rm 0-\kappa}$ is given by

\begin{equation}\label{eq:T_0_kappa_over_T_0_thr}
\frac{T_{\rm 0-\kappa}}{T_{\rm 0-thr}} =
         \begin{cases}\displaystyle{1-\frac{1}{\beta -2}\frac{(1+\kappa \,c_{\rm vir})^{\beta -2}-1}{\kappa \,c_{\rm vir}(1+\kappa \,c_{\rm vir})^{\beta -2}}} & \mbox{if $\beta \neq 2$,} \cr
                      \displaystyle{1-\frac{\ln(1+\kappa \,c_{\rm vir})}{\kappa \,c_{\rm vir}}} & \mbox{if $\beta = 2$.} \cr
         \end{cases}
\end{equation}

It can be shown that the ratio $T_{\rm 0-\kappa}/T_{\rm 0-thr}>0.9$ for $\kappa>10$ (i.e. $r_{\rm max}>10\,r_{\rm vir}$), in the typical range $4<c_{\rm vir}<10$ at $z=0$, for a range of virial masses $10^{12}$ M$_{\rm \odot} < M_{\rm vir} < 10^{16}$ M$_{\rm \odot}$, for any $2 \leq \beta \leq 3$.

In extended systems, gas quantities at large radii are never really zero but are equal to those of the intracluster medium or of the intercluster medium, depending on the type of system one considers. Assuming global HE, we can compute the central temperature when one does not impose $T_{\rm \infty}=0$, but imposes instead $T_{\rm \infty}=T_{\rm 0}/F$, where $F$ is some large number. In this case, the new central temperature $T_{\rm 0-F}$ is given by

\begin{equation}\label{eq:T_0_F_over_T_0_thr}
\frac{T_{\rm 0-F}}{T_{\rm 0-thr}} = \frac{F}{F-1}.
\end{equation}

The ratio $T_{\rm 0-F}/T_{\rm 0-thr}\lesssim 1.01$ for $F \gtrsim 100$ (i.e. $T_{\rm \infty} \lesssim T_{\rm 0}/100$), for any $1 < \beta \leq 3$. In clusters of galaxies, where $F>>1$, we can safely assume $T_{\rm \infty}=0$. Groups of galaxies, however, can have a much smaller $F$, and the assumption of a zero temperature at infinity cannot be used. Written in a different way: in clusters of galaxies, $T_{\rm \infty}$ is negligible with respect to $T_{\rm 0-thr}$ (i.e. $T_{\rm 0}=T_{\rm 0-thr}+T_{\rm \infty}\simeq T_{\rm 0-thr}$). In groups of galaxies, on the other hand, a non-zero $T_{\rm \infty}$ cannot be neglected, and we will see in Section \ref{sec:Dependence_on_the_gas_temperature_at_large_radii} its effects on the $M$--$T$ relation.

Considering quantities at infinity, like $T_{\rm \infty}$, is mathematically convenient but obviously not realistic. For example, the HE assumption may break down at $\sim r_{\rm vir}$, where there may also be a virial shock. In that case, a more complete approach might include the solution of the Rankine-Hugoniot jump conditions at the shock front, in order to obtain proper boundary conditions for the polytropic solution. A similar approach is required if one assumes the DM profile to have a sharp truncation radius at $r_{\rm vir}$ (e.g. Ostriker et al. 2005). However, note that while numerical simulations can show virial shocks at $\sim r_{\rm vir}$ along particular directions from the cluster centre (e.g. pointing to filaments), when the cluster profiles are spherically averaged the simulations do not show strong DM or gas pressure and density discontinuities at $r_{\rm vir}$ (e.g. Molnar et al. 2009). Moreover, given the uniqueness of the solution of the HE differential equation, once the Dirichlet boundary condition has been enforced, the imposition of $T(+\infty)=T_{\rm \infty}$ can be mapped to an equivalent boundary condition $T_{\rm \xi}$ at some finite radius $\xi\,r_{\rm vir}$, with $\xi \geq 0$:

\begin{equation}\label{eq:T_xi_NFW}
T_{\rm \xi}=T_{\rm \infty}\frac{1-(\Gamma-1)\Delta_{\rm gas}\left[1-\ln(1+\xi\,c_{\rm vir})/\xi \,c_{\rm vir}\right]/\Gamma}{1-(\Gamma-1)\Delta_{\rm gas}/\Gamma}
\end{equation}

\vspace{6pt}
\noindent for $\beta=2$ and

\begin{equation}\label{eq:T_xi_B10}
T_{\rm \xi}=T_{\rm \infty}\frac{1-\frac{\Gamma-1}{\Gamma}\Delta_{\rm gas}\left[1-\frac{1}{\beta-2}\frac{(1+\xi \,c_{\rm vir})^{\beta-2}-1}{(\xi \,c_{\rm vir})(1+\xi \,c_{\rm vir})^{\beta-2}}\right]}{1-(\Gamma-1)\Delta_{\rm gas}/\Gamma}
\end{equation}

\noindent for $\beta \neq 2$. These formulas will be useful in Section \ref{sec:Dependence_on_the_gas_temperature_at_large_radii}. It is also worth remembering that observational quantities like $M_{\rm 500}$ and $T_{\rm 500}$ (see Section \ref{sec:Average_temperature_and_the_issue_of_excision}) do not depend strongly on the exact solution at $\gtrsim r_{\rm vir}$, because of the low gas density and emissivity at large radii.

%SECTION 2.2
%SECTION 2.2
%SECTION 2.2
%SECTION 2.2
%SECTION 2.2

\subsection{Analysing the polytropic solution}\label{sec:Analysing_the_polytropic_solution}

In this section, we show the behaviour of $r_{\rm max}$ and $T_{\rm 0-thr}$ in a couple of different cases. In Figure \ref{fig:Dependence_on_z}, we show the dependence of $r_{\rm max}$ as a function of redshift for a cluster of virial mass $M_{\rm vir}=10^{14}$M$_{\rm \odot}$, with gas of polytropic index $\Gamma=6/5$ and central temperature $k_{\rm B}T_{\rm 0}=2.5$ keV, $f_{\rm gas}=f_{\rm b}$, and with DM described by an NFW profile and concentration given by D08. We also plot the virial radius and the threshold central temperature to allow for comparison. Notice how the threshold central temperature slowly increases by only a factor of $\sim$2 (from $k_{\rm B}T_{\rm 0-thr}\sim$2 to $\sim$5 keV) in the $z=0\,$--$\,3$ range\footnote{Note that, at high redshift, some of the assumptions of our model, including HE and isolation, are less valid.}: for a fixed virial mass, the strength of the gravitational potential only slowly increases with redshift. Finally, we also show the gas parameter $\Delta_{\rm gas}$, to compare it to $\Gamma/(\Gamma-1)$, and the temperature at infinity $T_{\rm \infty}$. The dashed, vertical line in all panels denotes the asymptote of the function $r_{\rm max}(z)$ and also denotes the threshold redshift $z_{\rm thr}$ at which $\Delta_{\rm gas}=\Gamma/(\Gamma-1)$, $T_{\rm 0}=T_{\rm 0-thr}$ and $T_{\rm \infty}=0$: below $z_{\rm thr}$, the solution is physically meaningful at all radii. Above $z_{\rm thr}$, the radius $r_{\rm max}\neq +\infty$ and decreases as redshift increases. Assuming HE at all radii, this threshold redshift is the only allowed redshift for this cluster, if we strictly impose $T_{\rm \infty}=0$, or is just a maximum redshift, if we allow for any non-negative $T_{\rm \infty}$.

In Figures \ref{fig:Dependence_on_z_varying_T} and \ref{fig:Dependence_on_z_varying_Gamma}, we show what changes when we vary the central temperature (by $\pm10$ per cent) or the polytropic index (by $\pm2$ per cent) of the gas, respectively. If we decrease the central temperature, the threshold redshift $z_{\rm thr}$ decreases, because the gas parameter $\Delta_{\rm gas}$ increases, and there is a \emph{minimum} central temperature (in this example $k_{\rm B}T_{\rm 0-min}\simeq1.8$ keV) under which a cluster with a given virial mass and polytropic index cannot exist in HE at all radii, at any redshift. Analogously, if we increase the polytropic index, the threshold redshift decreases, because the threshold central temperature increases, and there is a \emph{maximum} polytropic index (in this example $\Gamma_{\rm max} \simeq 13/10$) above which a cluster with a given virial mass and central temperature cannot exist in HE at all radii, at any redshift (see also Arieli \& Rephaeli 2003). The existence of a minimum central temperature and of a maximum polytropic index is due to the fact that the threshold central temperature at redshift $z=0$ is finite and is an increasing function of $\Gamma$. Notice that $\Gamma_{\rm max}$ increases, for a fixed virial mass, if we increase the central temperature. However, there is an additional constraint on the polytropic index, given by the condition of convective stability. The solution given in Equations \eqref{eq:polytropic_rho_NFW} and \eqref{eq:polytropic_rho_B10} is convectively unstable when $(\rho/P) dP/dr < \gamma \,d\rho /dr$, which translates to $\Gamma>\gamma$, where the adiabatic exponent $\gamma=5/3$, for a completely ionized, ideal, monatomic, non-degenerate gas.

On the other hand, if we increase $T_{\rm 0}$, the threshold redshift increases. The same happens if we decrease the polytropic index of the gas, because the threshold central temperature decreases. This trend can be seen also in another way. When we decrease the polytropic index towards its isothermal limit $\Gamma_{\rm lim}=1$, the relation between pressure and density can be written as $P(r)=[\rho(r)/\rho_{\rm 0}]T_{\rm 0}$. In this limiting case, Equation \eqref{eq:polytropic_rho_NFW} is no longer valid for $\beta=2$, and the solution to the HE equation, using once again the boundary condition $\rho(0) \equiv \rho_{\rm 0}$, becomes\footnote{In the $\beta \neq 2$ case, the isothermal solution is
\begin{equation}
\rho(r) = \rho_{\rm 0}\exp{\left[\Delta_{\rm gas}\left(\frac{1}{\beta -2}\frac{(1+r/r_{\rm s})^{\beta -2}-1}{(r/r_{\rm s})(1+r/r_{\rm s})^{\beta -2}}-1\right)\right]}>0. \nonumber
\end{equation}}
$\rho(r)=\rho_{\rm 0}\exp[-\Delta_{\rm gas}(1-\ln(1+r/r_{\rm s})/(r/r_{\rm s}))]$ (Makino, Sasaki \& Suto 1998). It is clear from this solution that the gas quantities are never negative at any radius, therefore there is no minimum central temperature requirement: $T_{\rm 0-thr}=0$.

In the example shown in Figure \ref{fig:Dependence_on_z}, $z_{\rm thr}=0.85$ for $\Gamma=6/5$. We will see in Section \ref{sec:Dependence_on_the_polytropic_index} that the polytropic index has typical values between 6/5 and 13/10. This means that the threshold redshift computed for $\Gamma=6/5$ is indeed a maximum value for the example shown, for any reasonable value of the polytropic index. If we could observe with enough accuracy the central temperature of the cluster gas, we could then obtain some useful information. Using the numbers above, if we for example observed a cluster of virial mass $M_{\rm vir} = 10^{14}$ M$_{\rm \odot}$ with a gas with polytropic index $\Gamma \ge 6/5$ and central temperature $k_{\rm B}T_{\rm 0}=2.5$ keV to be at redshift $z \ge 1$, then we could safely assume that the gas is \emph{not} in HE at all radii. Alternatively, if we for example observed a cluster with the same virial mass and central temperature at redshift $z \simeq 0.5$ and knew by some means that the gas is in global HE, then we could safely assume that it has a $\Gamma>6/5$. Unfortunately, accurate observations of $T_{\rm 0}$ are very difficult if not in many cases impossible, due to resolution limits of current X-ray telescopes. However, the same reasoning translates to the concept of average temperature, defined in Section \ref{sec:Average_temperature_and_the_issue_of_excision}. This will be evident in Section \ref{sec:The_M_T_Relation}, where we show the dependence of the $M$--$T$ relation on several parameters, including the polytropic index and redshift.

\begin{figure}
\centering
\vspace{5pt}
\includegraphics[width=0.7\columnwidth,angle=90]{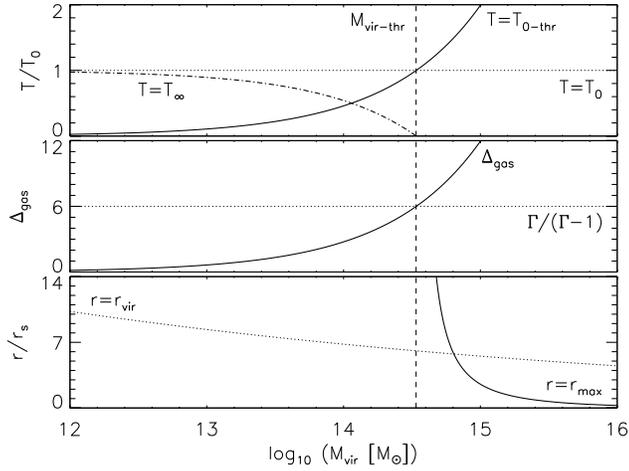}
\vspace{10pt}
\caption{Dependence on virial mass of several parameters, for a cluster at redshift $z=0$, composed of NFW DM and gas with polytropic index $\Gamma=6/5$ and central temperature $k_{\rm B}T_{\rm 0}=4$ keV. {\bf Top panel}: we plot the temperature of the cluster $T_{\rm 0}$ (\emph{dotted line}), its central threshold temperature $T_{\rm 0-thr}$ (\emph{solid line}) and its temperature at infinity $T_{\rm \infty}$ (\emph{dot-dashed line}). Notice how $T_{\rm 0-thr}$ depends strongly on virial mass, changing by more than one order of magnitude in the $M_{\rm vir}=10^{12}\,$--$\,10^{16}$ M$_{\rm \odot}$ range. {\bf Middle panel}: we plot the gas parameter $\Delta_{\rm gas}$ (\emph{solid line}) in relation to the ratio $\Gamma/(\Gamma-1)$ (\emph{dotted line}). {\bf Bottom panel}: we plot the virial radius $r_{\rm vir}$ (\emph{dotted line}) and the maximum radius $r_{\rm max}$ (\emph{solid line}) of the cluster. {\bf All panels}: the \emph{dashed, vertical line} denotes the threshold virial mass $M_{\rm vir-thr}$ at which $T_{\rm 0}=T_{\rm 0-thr}$, $T_{\rm \infty}=0$, $\Delta_{\rm gas}=\Gamma/(\Gamma-1)$ and the function $r_{\rm max}(M_{\rm vir})$ has its vertical asymptote. When we impose $T_{\rm \infty}=0$, $M_{\rm vir}=M_{\rm vir-thr}$ is the only mass that the cluster can have to exist in HE at all radii. If we instead allow for any $T_{\rm \infty} \geq 0$, then the cluster can exist in HE at all radii with a virial mass $M_{\rm vir}=0\,$--$\,M_{\rm vir-thr}$.}
\label{fig:Dependence_on_M}
\end{figure}

We now study the dependence of the HE-polytropic solution on the mass of the system. In Figure \ref{fig:Dependence_on_M}, we show the same parameters of Figure \ref{fig:Dependence_on_z}, this time as a function of virial mass, for a cluster at redshift $z=0$ with NFW DM, $f_{\rm gas}=f_{\rm b}$, and gas with polytropic index $\Gamma=6/5$ and central temperature $k_{\rm B}T_{\rm 0}=4$ keV. Notice how the threshold central temperature increases significantly as the virial mass increases: for a fixed redshift, the strength of the gravitational potential quickly increases with virial mass. As a consequence, $r_{\rm max}$ strongly depends on virial mass, changing from $r_{\rm max} \sim r_{\rm vir}$ to $+\infty$ in a very small mass range.

\begin{figure}
\centering
\vspace{5pt}
\includegraphics[width=0.7\columnwidth,angle=90]{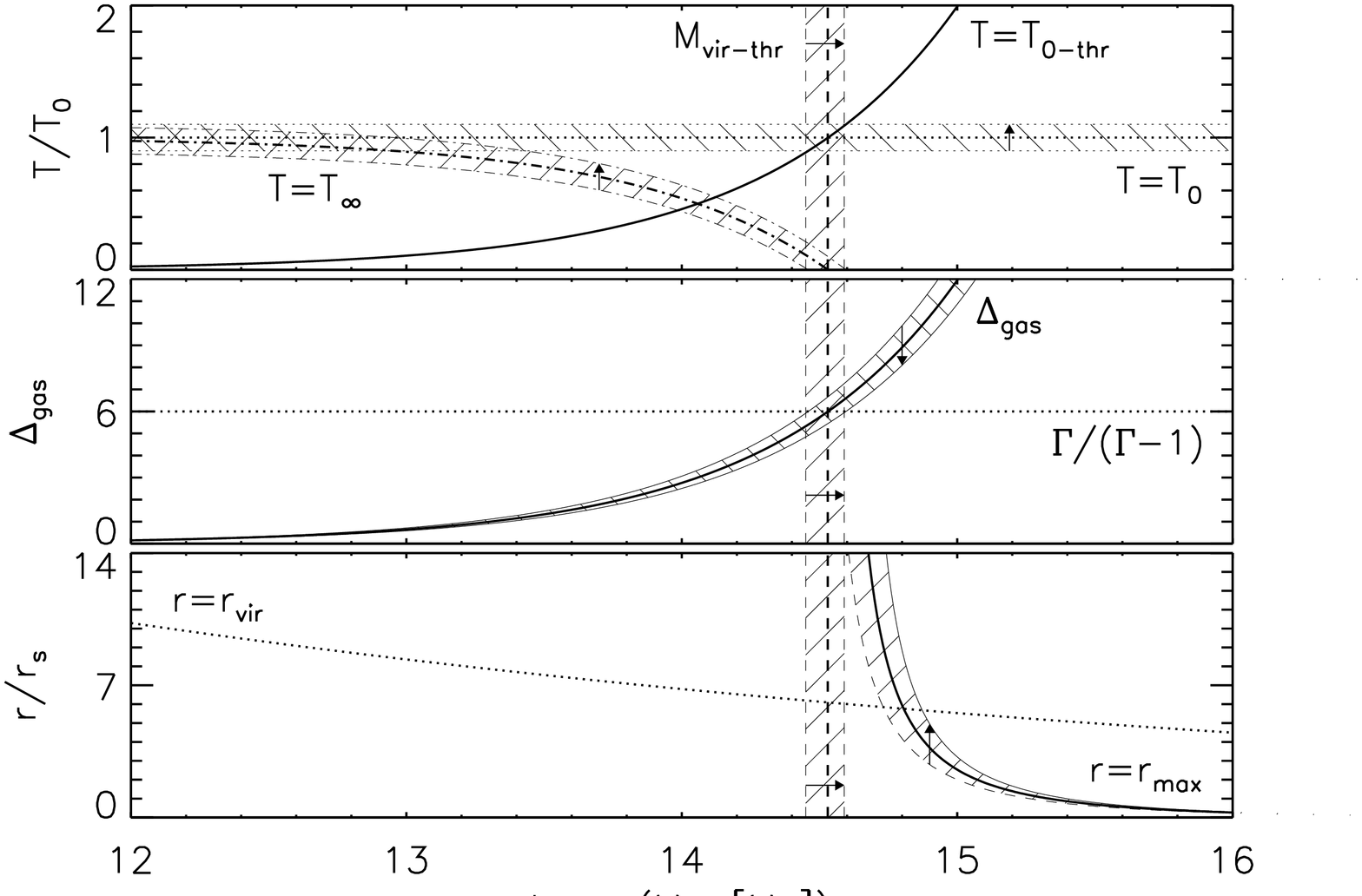}
\vspace{0pt}
\caption{Dependence of the parameters of the cluster described in Figure \ref{fig:Dependence_on_M} on its gas central temperature. The thick lines are the same lines shown in Figure \ref{fig:Dependence_on_M}. The arrows and the bands show what changes when we vary the central temperature from $0.9\,T_{\rm 0}$ to $1.1\,T_{\rm 0}$. When increasing the central temperature, $\Delta_{\rm gas}$ decreases, $T_{\rm \infty}$ and $r_{\rm max}$ increase and the threshold virial mass $M_{\rm vir-thr}$ increases.}
\label{fig:Dependence_on_M_varying_T}
\end{figure}

\begin{figure}
\centering
\vspace{5pt}
\includegraphics[width=0.7\columnwidth,angle=90]{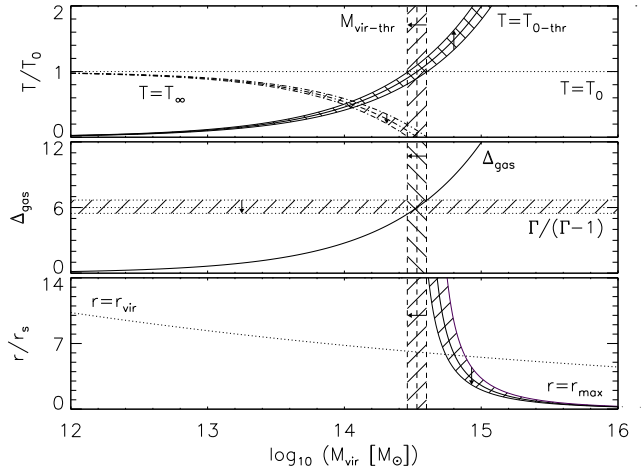}
\vspace{0pt}
\caption{Dependence of the parameters of the cluster described in Figure \ref{fig:Dependence_on_M} on its gas polytropic index. The thick lines are the same lines shown in Figure \ref{fig:Dependence_on_M}. The arrows and the bands show what changes when we vary the polytropic index from $0.98\,\Gamma$ to $1.02\,\Gamma$. When increasing the polytropic index, $T_{\rm 0-thr}$ increases, $T_{\rm \infty}$ and $r_{\rm max}$ decrease and the threshold virial mass $M_{\rm vir-thr}$ decreases.}
\label{fig:Dependence_on_M_varying_Gamma}
\end{figure}

In Figures \ref{fig:Dependence_on_M_varying_T} and \ref{fig:Dependence_on_M_varying_Gamma}, we show what changes when we vary the central temperature (by $\pm10$ per cent) or the polytropic index (by $\pm2$ per cent) of the gas, respectively. The effect is similar to what shown in Figures \ref{fig:Dependence_on_z_varying_T} and \ref{fig:Dependence_on_z_varying_Gamma}, except for the fact that there is no $T_{\rm 0-min}$ or $\Gamma_{\rm max}$: for every small central temperature or large polytropic index, there is always a small enough virial mass for which the DM gravitational potential is not strong enough to bind the gas in a finite region. Remember, however, that $\Gamma<\gamma$, because of the convective stability requirement.

%SECTION 2.3
%SECTION 2.3
%SECTION 2.3
%SECTION 2.3
%SECTION 2.3

\subsection{Comparison to previous studies}\label{sec:Comparison_to_previous_studies}

In Section \ref{sec:Formulation}, we describe the structure of gas in clusters and groups of galaxies of known virial mass and redshift, by assuming HE at all radii between an ideal polytropic gas and DM. If the DM potential is known, the HE-polytropic solution is completely described once we fix its three free parameters: $\Gamma$, $\rho_{\rm 0}$ and $T_{\rm 0}$. There are several ways to ``close the problem'', which are all mathematically equivalent. Here we briefly summarise the assumptions we made in Section \ref{sec:Formulation} and then contrast them with those made in prior work. This will help put into better context our results in Section \ref{sec:Scaling_relations_and_comparison_to_observations} and those of other authors.

For a given polytropic index, we obtain a relation between the gravitational potential of a system and a range of permitted central temperatures of its gas, by simply requiring (i) spherical symmetry, (ii) HE between DM (described by a generalised NFW profile) and an ideal polytropic gas at all radii, and by imposing (iii) that the temperature at large radii is non-negative (see Equation \ref{eq:T_0_thr} and related discussion).

KS01 (see also KS02) presented a similar description for the intracluster medium in clusters, by also assuming (i) and (ii), and additionally requiring (iv) that the gas density profile is equal to that of DM at large radii, (v) the DM large-radii logarithmic slope $\zeta=-3$ and (vi) the concentration is known, for a given mass and redshift.

A06 (see also Ascasibar et al. 2003) also studied cluster models of polytropic gas, by once again requiring (i), (ii), (v) and (vi), additionally (vii) taking the polytropic index as a given (from numerical simulations and observations), and imposing (viii) that the temperature at infinity is always zero, regardless of the system's mass or environment.

Ostriker et al. (2005) (but see also Bode et al. 2007, 2009) studied similar scenarios, by also assuming (i), (ii), (v), (vi) and (vii). Moreover, (ix) they constrained the central temperature by matching the external surface pressure to the momentum flux from the infalling gas at the virial radius.

%SECTION 2.4
%SECTION 2.4
%SECTION 2.4
%SECTION 2.4
%SECTION 2.4

\subsection{Average temperature and the issue of excision}\label{sec:Average_temperature_and_the_issue_of_excision}

In this section, we connect the results given in Section \ref{sec:Formulation} to observations, by means of defining the average temperature. As shown in the previous sections, for a given polytropic index, we are able to obtain a relation between the gravitational potential of a galaxy group or cluster and its gas central temperature, once a boundary condition (e.g. $T_{\rm \infty}$) has been fixed. Naturally, the central temperature is not an easy observable, because of resolution limits, but it is possible to apply the same reasoning and results to the concept of average temperature. We computed the average temperature following the notation of KS01. In general, one can write

\begin{equation}
T_{\rm av} = \frac{\int_{\rm r_{\rm 0}}^{r_{\rm 1}}w(r)T(r)4\pi r^2dr}{\int_{\rm r_{\rm 0}}^{r_{\rm 1}}w(r)4\pi r^2dr},
\end{equation}

\noindent where the excision radius $r_{\rm 0}$ is either zero or a fraction of an overdensity radius (e.g. $r_{\rm vir}$, $r_{\rm 200}$, $r_{\rm 500}$, or other), the truncation radius $r_{\rm 1}$ is some other fraction of, or equal to, an overdensity radius, and $w(r)$ is the weight function. If the weight function $w=1$, $T_{\rm av}=T_{\rm vwa}$ is the volume-weighted average temperature; if $w=\rho$, $T_{\rm av}=T_{\rm mwa}$ is the mass-weighted average temperature; if $w=n_{\rm i}n_{\rm e} \Lambda_{\rm N}$, where $n_{\rm e}$ is the number density of electrons per unit volume, $n_{\rm i}$ is the number density of ions per unit volume and $\Lambda_{\rm N}$ is the cooling function, $T_{\rm av}=T_{\rm ewa}$ is the emission-weighted average temperature. Although some authors use the mass-weighted average temperature, most observers use the emission-weighted average temperature instead, usually approximating the cooling function as $\Lambda_{\rm N} \propto \rho^2 T^{1/2}$. Accordingly, we also use the emission-weighted temperature here. However, since we also investigate the group regime, where temperatures are lower, we use a more general version of the cooling function, described in detail in Appendix \ref{sec:Cooling_Function}. When using both definitions of the cooling function, we find the results to be basically identical at high temperature (as expected, since cooling is dominated by Bremsstrahlung emission) and only slightly different at low temperatures.

The choice of $r_{\rm 0}$ and $r_{\rm 1}$ is an important one. The truncation radius $r_{\rm 1}$ is usually determined by observational constraints. At large radii, both gas temperature and density are very low, therefore very difficult to quantify accurately. However, exactly because temperature and density are very low, $T_{\rm ewa}$ does not change appreciably for radii greater than $r_{\rm 500}$, except for the highest mass systems, and does not change at all for radii greater than $r_{\rm vir}$. Even when future telescopes (e.g. eROSITA; Predehl et al. 2007, 2010) will permit us to study the outer regions with better accuracy, a larger $r_{\rm 1}$ should not significantly change the average temperature.

The choice of the excision radius $r_{\rm 0}$ is less straightforward. Most studies justify excising the central region because it is significantly affected by radiative cooling and by other physics of which we do not know the details, e.g. feedback. The positive result of excision is the large decrease in scatter in the X-ray scaling relations, at all masses (e.g. Markevitch 1998), which is highly useful for cosmology measurements. However, excision also throws out potentially important physical information relevant to understanding in detail the formation of the cluster, such as the effects of cooling, feedback, concentration, etc. Therefore, it is useful to study how the choice of $r_{\rm 0}$ affects the results. In Section \ref{sec:Dependence_on_the_mass-concentration_relation}, we will see an example of how significant excision can be.

Cognizant of all these considerations, in this paper we choose to use the integration limits of V09 (see also Kravtsov, Vikhlinin \& Nagai 2006) and define the average temperature as\footnote{When computing average temperatures within other overdensity radii, we keep the excision radius the same. For example, the integration limits for the computation of $T_{\rm vir}$ are $r_{\rm 0}=0.15\,r_{\rm 500}$ and $r_{\rm 1}=r_{\rm vir}$.}

\begin{equation}\label{eq:T_average}
T_{\rm 500} = \frac{\int_{\rm 0.15\,r_{\rm 500}}^{r_{\rm 500}}n_{\rm e}n_{\rm i}\Lambda_{\rm N}T(r)4\pi r^2dr}{\int_{\rm 0.15\,r_{\rm 500}}^{r_{\rm 500}}n_{\rm e}n_{\rm i}\Lambda_{\rm N}4\pi r^2dr}.
\end{equation}

When studying the effect of excision, we compute the average temperatures $T_{\rm 500-no-exc}$ and $T_{\rm vir-no-exc}$, simply by changing the lower integration limit to $r_{\rm 0}=0$.

The average temperature described here will be used in Section \ref{sec:The_M_T_Relation}, where we calculate the $M$--$T$ relation for groups and clusters of galaxies and compare it to recent X-ray observations.

%SECTION 3
%SECTION 3
%SECTION 3
%SECTION 3
%SECTION 3

\section{Scaling relations and comparison to observations}\label{sec:Scaling_relations_and_comparison_to_observations}

%SECTION 3.1
%SECTION 3.1
%SECTION 3.1
%SECTION 3.1
%SECTION 3.1

\subsection{The $M$--$T$ relation}\label{sec:The_M_T_Relation}

In Section \ref{sec:Formulation}, we effectively obtained a relation between the temperature profile and the mass of clusters and groups of galaxies, once the temperature at large radii was fixed. For a proper comparison to observations, we defined the average temperature in Section \ref{sec:Average_temperature_and_the_issue_of_excision}. Using this quantity, we can then construct an $M$--$T$ relation that may be compared to observations.

In our model, the resulting $M$--$T$ relation is a function of five parameters: the polytropic index $\Gamma$, the temperature at large radii $T_{\rm \xi}$ (or, equivalently, $T_{\rm \infty}$), the DM logarithmic slope at large radii $\zeta$ (or, equivalently, $\beta$), the concentration $c_{\rm vir}$ and the redshift $z$. To develop intuition for how these parameters affect the $M$--$T$ relation, in the following sub-sections we will fix four of these quantities and study the dependence of the $M$--$T$ relation on the remaining fifth parameter. Comparing the predicted $M$--$T$ relation to that from X-ray observations, which have improved significantly in recent years, leads to constraints on the allowable ranges of parameter values. A thorough examination of the allowed range would require a simultaneous variation of all parameters at once, to allow for possible degeneracies. Such a study is beyond the scope of this paper, although where we can, we will note the more obvious degeneracies.

Before we embark on our study, some considerations are in order. We determine the temperature of a galaxy group or cluster by solving the thermally-supported HE equation. However, it is clear from simulations (e.g. Nagai, Vikhlinin \& Kravtsov 2007) that random gas motions and rotation can contribute up to 10 per cent of the pressure support, leading to an over-estimation of the gas temperature needed for HE. On the other hand, the mass computed from X-ray observations is the thermally-supported HE mass (e.g. Fabricant, Lecar \& Gorenstein 1980), resulting in an under-estimation of the true mass within $r_{\rm 500}$ of up to 10 per cent, for relaxed systems (e.g. Lau, Kravtsov \& Nagai 2009; see also Mahdavi et al. 2008). Since the $M$--$T$ relation is a power law, these two effects cancel each other and the comparison between our model and observations is meaningful (see also KS01). The same reasoning applies to any other form of non-thermal support, including cosmic rays and/or magnetic fields (see e.g. Parrish et al. 2011).

More importantly, in our model we assume a complete polytropic relation (i.e. the polytropic index is the same at all radii). Recent numerical studies (e.g. Shaw et al. 2010; Battaglia et al. 2011a,b) show instead a radial dependence. We note, however, that the variation of $\Gamma$ with radius is not very large, especially in the range $0.15\, r_{\rm 500}\,$--$\,r_{\rm 500}$, where we calculate the average temperature (see Section \ref{sec:Average_temperature_and_the_issue_of_excision}). These same studies (Battaglia et al. 2011b) also show a slight dependence of $\Gamma$ on redshift, and we will address this possibility in Section \ref{sec:Dependence_on_redshift}.

%SECTION 3.1.1
%SECTION 3.1.1
%SECTION 3.1.1
%SECTION 3.1.1
%SECTION 3.1.1

\subsubsection{Dependence on the polytropic index}\label{sec:Dependence_on_the_polytropic_index}

\begin{figure}
\centering
\vspace{5pt}
\includegraphics[width=0.72\columnwidth,angle=90]{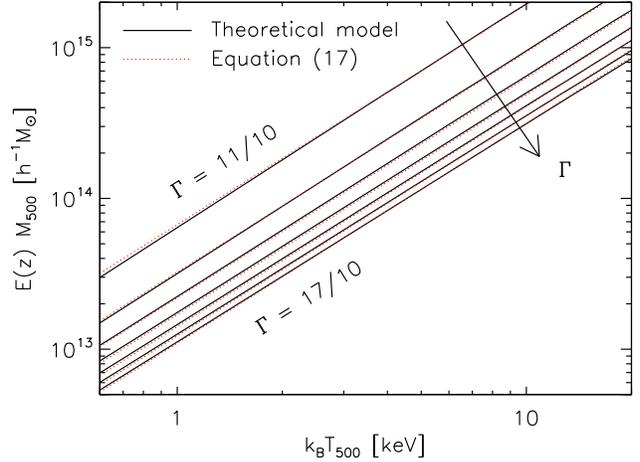}
\vspace{5pt}
\caption{Dependence of the theoretical $M$--$T$ relation on the polytropic index, for systems at redshift $z=0$ ($E(z)=1$), with polytropic gas in HE with NFW DM ($\zeta = -3$) at all radii, concentration given by D08 and $T_{\rm \infty}=0$. The relation can always be well approximated by a single power law. We plot the $M$--$T$ relation for seven evenly-spaced values of $11/10 \leq \Gamma \leq 17/10$. The \emph{black, solid lines} are the results from our theoretical model, whereas the \emph{red, dotted lines} show the empirical approximation from Equation \eqref{eq:MT_relation_Gamma}. Notice how the polytropic index has a significant effect only on the normalisation. For a fixed mass, the increase of $\Gamma$ results in a higher average temperature. In this and in the following figures, we assume our model scales self-similarly with redshift. We will see in Section \ref{sec:Dependence_on_redshift} that this assumption is well justified.}
\label{fig:MT_Relation_with_Gamma}
\end{figure}

In this section, we consider systems at redshift $z=0$ ($E(z)=1$), described by an NFW DM profile ($\zeta = -3$), with concentration given by D08 and a polytropic gas with $T_{\rm \infty}=0$, and study the dependence of the $M$--$T$ relation on the polytropic index $\Gamma$.

Figure \ref{fig:MT_Relation_with_Gamma} shows this dependence, when assuming that systems at all temperatures share the same value of $\Gamma$. We first notice that the relation can be very well approximated by a single power law, over more than three orders of magnitude in both mass and temperature (not shown in the figure). Moreover, the slope of such function does not notably change with $\Gamma$: the polytropic index has a significant effect only on the normalisation. With all these considerations, we can then approximate the local $M$--$T$ relation with the fitting\footnote{For this and some other fits in the paper, we performed a non-linear least-squares fitting with the IDL routine MPFIT (Markwardt 2009). Curiously enough, if we plot $M_{\rm vir}$ versus $T_{\rm 500}$, we obtain a relation with a self-similar slope: $M_{\rm vir}=M_{\rm 0}[\Gamma/(\Gamma-1)]^{6/5}(T_{\rm 500}/T_{\rm 0})^{3/2}$, where $M_{\rm 0}=7.09 \times 10^{13} \, h^{-1}$M$_{\rm \odot}$ and $k_{\rm B}T_{\rm 0}=5$ keV.} formula

\begin{equation}\label{eq:MT_relation_Gamma}
E(z) M_{\rm 500}=M_{\rm 0}\left(\frac{\Gamma}{\Gamma-1}\right)^{\epsilon}\left(\frac{T_{\rm 500}}{T_{\rm 0}}\right)^{\alpha},
\end{equation}

\noindent where $\epsilon=6/5$, $\alpha=1.46$, $M_{\rm 0}=3.9545 \times 10^{13} \, h^{-1}$M$_{\rm \odot}$, $k_{\rm B}T_{\rm 0}=5$ keV and $k_{\rm B}T_{\rm 500}$ is in keV. This approximation closely reproduces the results of the model for typical values of $11/10\leq\Gamma \leq 17/10$. In this and in the following sections, we assume our model scales self-similarly with redshift. We will see in Section \ref{sec:Dependence_on_redshift} that this assumption is well justified.

The effect of varying the polytropic index is consistent with what was already shown in Figures \ref{fig:Dependence_on_z_varying_Gamma} and \ref{fig:Dependence_on_M_varying_Gamma}: for a given virial mass, when $\Gamma$ increases, the threshold central temperature increases, and so does the average temperature. When $\Gamma$ decreases, the average temperature quickly approaches zero proportionally to $\sim$$[(\Gamma-1)/\Gamma]^{4/5}$, as the polytropic relation tends to its isothermal limit (i.e. $\Gamma=1$). Notice that, because of the dependence on the $\Gamma/(\Gamma -1)$ term, it is much easier to discriminate between \emph{lower} values of the polytropic index than between \emph{higher} values.

\begin{figure}
\centering
\vspace{5pt}
\includegraphics[width=0.72\columnwidth,angle=90]{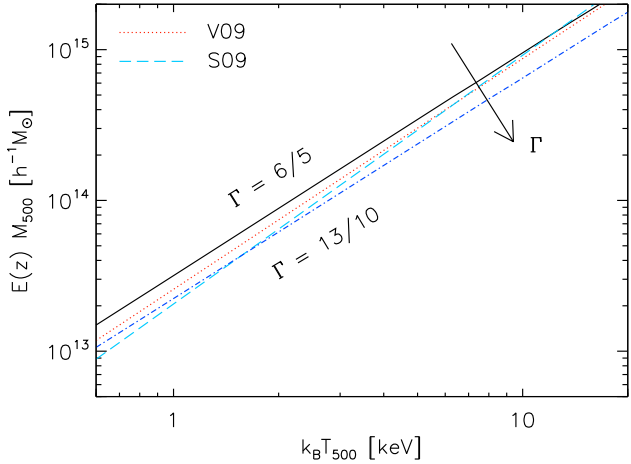}
\vspace{5pt}
\caption{Constraints on the polytropic index, for systems at redshift $z=0$ ($E(z)=1$), with polytropic gas in HE with NFW DM ($\zeta = -3$) at all radii, concentration given by D08 and $T_{\rm \infty}=0$, via comparison between our theoretical model and observations by V09 and S09 (\emph{red, dotted} and \emph{turquoise, long-dashed line}, respectively). We plot the theoretical $M$--$T$ relation for $\Gamma=6/5$ (\emph{upper, black, solid line}) and $\Gamma=13/10$ (\emph{lower, blue, dot-dashed line}). A model with a mass-independent $\Gamma$ with a value between 1.22 and 1.24 is consistent with the 1-$\sigma$ errors of V09. No model with a constant $\Gamma$ is consistent with S09. For consistency with S09, with all other parameters fixed, we need to assume a mass-dependent polytropic index (see Equation \ref{eq:Gamma_M_relation}).}
\label{fig:MT_Relation_with_Gamma_from_observations}
\end{figure}

We now compare the theoretical $M$--$T$ relation to recent X-ray observations of galaxy clusters and groups. The red, dot-dashed line in Figure \ref{fig:MT_Relation_with_Gamma_from_observations} shows the best-fitting $M$--$T$ relation calculated by V09 for a sample of seventeen, low-redshift, relaxed clusters: $E(z)M_{\rm 500} = (3.02 \pm 0.11) \times 10^{14} h^{-1}$M$_{\rm \odot} (k_{\rm B}T_{\rm 500}/5$ keV$)^{1.53 \pm 0.08}$, valid over the temperature range $k_{\rm B}T\sim 1.5\,$--$\,12$ keV. Since we do not have access to the exact measurement and error analysis of V09, we cannot make a definitive statement on the range of $\Gamma$ consistent with V09. Assuming zero covariance between the magnitude and the slope of the published $M$--$T$ relation, we find that a model with a mass-independent polytropic index $1.22 \lesssim \Gamma \lesssim 1.24$ is consistent with the data. Without this assumption, we conservatively estimate $6/5 \lesssim \Gamma \lesssim 13/10$. Such values of $\Gamma$ are also consistent with the results obtained by both observations and simulations (e.g. Bode et al. 2009 and references therein).

Observations of low-mass systems show increasing scatter and a steepening of the $M$--$T$ relation. It is not clear if the $M$--$T$ relation for all (low- and high-mass) systems should be described by a single power law (only steeper than the cluster-only case), or if there is a real break in the relation at low masses. The existence of such break is still under debate (e.g. Eckmiller, Hudson \& Reiprich 2011). Several authors (e.g. Nevalainen, Markevitch \& Forman 2000; Finoguenov, Reiprich \& B{\"o}hringer 2001; Sanderson et al. 2003) suggest its presence at around $k_{\rm B}T \sim 1$ keV, but other authors (e.g. S09) do not find conclusive evidence for it. We will consider the possibility of a broken power law in Section \ref{sec:Dependence_on_the_gas_temperature_at_large_radii} and focus here on comparing our model to a single, steeper power law. The turquoise, long-dashed line in Figure \ref{fig:MT_Relation_with_Gamma_from_observations} shows the best-fitting $M$--$T$ relation calculated by S09 for a sample of thirty-seven, low-redshift, relaxed groups and clusters (including fourteen clusters from V09): $E(z)M_{\rm 500} = (1.26 \pm 0.07) \times 10^{14} h^{-1}$M$_{\rm \odot} (k_{\rm B}T_{\rm 500}/3$ keV$)^{1.65 \pm 0.04}$, valid over the temperature range $k_{\rm B}T\sim 0.7\,$--$\,12$ keV. In this case, our model with a constant polytropic index is clearly not consistent with the data.

A possible explanation for this discrepancy is that the polytropic index is not the same for all systems, but instead depends on the mass of the system. By direct comparison of the theoretical model to the best-fitting $M$--$T$ relation by S09, we obtain an empirical relation between $\Gamma$ and the virial mass of local groups and clusters of galaxies,

\begin{equation}\label{eq:Gamma_M_relation}
\Gamma(M_{\rm vir}) = 1.305 - 0.073 \log_{\rm 10}\left(\frac{M_{\rm vir}}{10^{14} \hbox{M}_{\rm \odot}}\right),
\end{equation}

\noindent which is valid over the combined V09 and S09 temperature ranges -- $k_{\rm B}T\sim 0.7\,$--$\,12$ keV -- and where $M_{\rm vir}$ is in M$_{\rm \odot}$.

The polytropic index slowly decreases with mass. For a proper comparison with previous literature, we convert this equation to a relation between $\Gamma$ and concentration, by using Equation \eqref{eq:c_vir_M_vir_relation}:

\begin{equation}\label{eq:Gamma_c_relation}
\Gamma(c_{\rm vir}) = 1.305 + 0.816 \log_{\rm 10}\left(\frac{c_{\rm vir}}{6.8}\right),
\end{equation}

\noindent where 6.8 is the concentration of a local cluster of virial mass equal to $10^{14}$M$_{\rm \odot}$, according to D08.

KS01 (see also KS02) and A06 each obtain a relation between $\Gamma$ and $c_{\rm vir}$. Although in all three relations the polytropic index increases with concentration, the slope and normalisation are quite different. A06, by using data from their own simulations, find a polytropic index which is almost constant with concentration: $\Gamma = 1.145+0.0036 \,c_{\rm vir}$. KS02, by requiring that the gas profile at large radii is equal to that of DM, find a slightly steeper slope than A06, obtaining the relation $\Gamma = 1.137+0.0894 \ln (c_{\rm vir}/5) - 0.00368 (c_{\rm vir}-5)$, valid\footnote{KS01 provide a simpler relation, $\Gamma = 1.15 + 0.01(c_{\rm vir}-6.5)$, valid only for the temperature range we are considering in this paper.} for $1<c_{\rm vir}<25$. In both cases, the polytropic index in the combined temperature range of S09 and V09 ($k_{\rm B}T\sim 0.7\,$--$\,12$ keV, $c_{\rm vir}\sim 5\,$--$\,7.5$) is almost constant and close to 1.15, a value slightly lower than that of current observations and numerical simulations (e.g. Bode et al. 2009 and references therein). In that same temperature range, our relation is steeper than both A06 and KS02 (and KS01), with the polytropic index varying from $\sim$13/10 in the group regime to $\sim$6/5 in the cluster regime. If we assume our theoretical model and the observations we use as comparison (S09) to be correct, one implication of this discrepancy with KS01 and KS02 is the following: although qualitatively reasonable, the assumption that the slopes of gas and DM profiles at large radii are equal cannot be used to construct an accurate model of gas in groups and clusters of galaxies.

It is not immediately clear why the polytropic index should depend on the mass (or, equivalently, concentration) of the system. Following Bertschinger (1985), $\Gamma$ is determined by events like cluster collapse and relaxation, which might depend on mass scale and redshift, so that such a result may not be surprising. Also, feedback efficiency might be a strong function of halo mass and cause a mass-dependent polytropic index. However, we remind the reader that this constraint on $\Gamma$ was obtained after fixing all other free parameters. In Section \ref{sec:Dependence_on_the_gas_temperature_at_large_radii}, we present a different explanation for the steepening/breaking of the $M$--$T$ relation, wherein a mass-dependent polytropic index is not necessary and the shape of the power law changes with the choice of the temperature at large radii. Of course, both these causes could be at work at the same time.

%SECTION 3.1.2
%SECTION 3.1.2
%SECTION 3.1.2
%SECTION 3.1.2
%SECTION 3.1.2

\subsubsection{Dependence on the gas temperature at large radii}\label{sec:Dependence_on_the_gas_temperature_at_large_radii}

\begin{figure}
\centering
\vspace{5pt}
\includegraphics[width=0.72\columnwidth,angle=90]{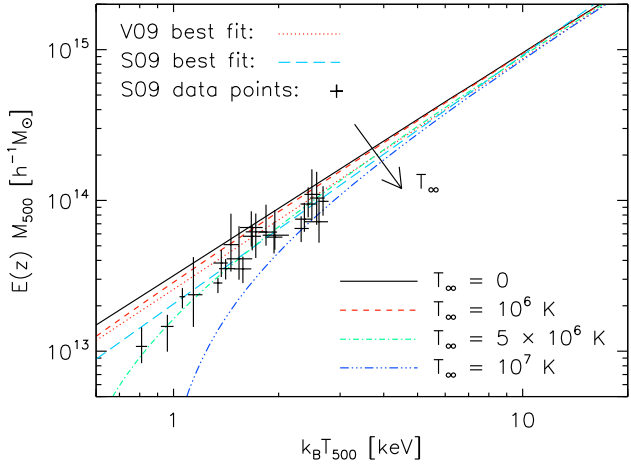}
\vspace{5pt}
\caption{Constraints on the gas temperature at large radii, for systems at redshift $z=0$ ($E(z)=1$), with polytropic gas in HE with NFW DM ($\zeta = -3$) at all radii, concentration given by D08 and $\Gamma=6/5$, via comparison between our theoretical model and observations by V09 (\emph{red, dotted line}) and by S09 (\emph{turquoise, long-dashed line} and \emph{data points} with 1-$\sigma$ errors). We plot the theoretical $M$--$T$ relation for $T_{\rm \infty}=0$ (\emph{black, solid line}), $T_{\rm \infty}=10^6$K (\emph{red, short-dashed line}), $T_{\rm \infty}=5\times 10^6$K (\emph{green, dot-dashed line}) and $T_{\rm \infty}=10^7$K (\emph{blue, triple-dot-dashed line}). A model with $T_{\rm \infty}=0$ is consistent with V09 but not with S09. For consistency with S09, with all other parameters fixed, we need to assume $10^6$ K $\lesssim T_{\rm \infty} \lesssim 10^7$ K, in agreement with numerical simulations and observations of the warm-hot intergalactic medium.} 
\label{fig:MT_Relation_with_Tinf}
\end{figure}

In this section, we consider systems at redshift $z=0$ ($E(z)=1$), described by an NFW DM profile ($\zeta = -3$), with concentration given by D08 and a polytropic gas with $\Gamma=6/5$, and study the dependence of the $M$--$T$ relation on the temperature at large radii.

In Section \ref{sec:Polytropic_gas_in_galaxy_groups_and_clusters}, after imposing HE at all radii, we found a range of permitted values for the central gas temperature, $T_{\rm 0-min}\,$--$\,T_{\rm 0-max}$, where $T_{\rm 0-min}=T_{\rm 0-thr}$ is constrained by requiring $T_{\rm \infty}=0$ and $T_{\rm 0-max}$ is given by some different boundary condition at large radii ($T_{\rm \infty} \neq 0$). Here, we explore this range of permitted temperatures further, by studying the dependence of the $M$--$T$ relation on $T_{\rm \xi}$, the gas temperature at a given radius $\xi\,r_{\rm vir}$. As demonstrated in Section \ref{sec:Formulation}, there is a one--to--one relation between $T_{\rm \xi}$ and $T_{\rm \infty}$, for any given value of $\xi$. Therefore, we can perform all calculations using $T_{\rm \infty}$ as the boundary condition for simplicity, and then translate the results back to $T_{\rm \xi}$.

As shown in Figure \ref{fig:MT_Relation_with_Tinf}, when imposing $T_{\rm \infty} \neq 0$, the $M$--$T$ relation ceases to be a single power law. Although there is basically no difference at high masses (for any reasonable value of $T_{\rm \infty}$), the $M$--$T$ relation steepens at low masses, reproducing an effective break. The larger $T_{\rm \infty}$ is, the higher is the mass (and temperature) at which the turn-off should be observed. A steepening/breaking in the $M$--$T$ relation has already been mentioned in Section \ref{sec:Dependence_on_the_polytropic_index}, as a possible feature of the scaling relation of low-mass systems (e.g. Eckmiller et al. 2011). In Section \ref{sec:Dependence_on_the_polytropic_index}, we considered the single-power-law description and proposed that a mass-dependent polytropic index could explain the $M$--$T$ relation that fits both low- and high-mass quantities (e.g. S09). Here, we consider the broken-power-law description and explain the existence of the break with a non-zero $T_{\rm \infty}$. We note, however, that the explanations we give are independent of the description of the $M$--$T$ relation; a mass-dependent polytropic index can reproduce a broken power law, whereas a non-zero $T_{\rm \infty}$ can also explain a single, steep power law. Likely, these two causes are concurrent.

We propose here two interpretations for the temperature at large radii. If we assume $T_\infty$ to be the same for every system -- as one would expect, if one thinks of all isolated groups and clusters to be embedded in the same intercluster medium -- we can constrain the lower and upper limits of the temperature of the external medium, since we actually observe a possible turn-off at around $k_{\rm B}T \sim 1$ keV (e.g. Sanderson et al. 2003; see also Dos Santos \& Dor{\'e} 2002). In order to match the observations of galaxy groups, $T_\infty$ must be between $10^6$ K and $10^7$ K. This in turn results in different values of $T_{\rm \xi}$ for different values of virial mass (and of $\xi$), from using Equation \eqref{eq:T_xi_NFW} (or Equation \ref{eq:T_xi_B10}, if $\beta \neq 2$). For example, if we set $\xi=5$ (i.e. if we are interested in the temperature at five virial radii) and $T_{\rm \infty}=5 \times 10^6$ K, we obtain $T_{\rm \xi} \simeq 0.5$, $\simeq 0.8$ and $\simeq 2$ keV, for local systems with $M_{\rm 500}=10^{13}$, $10^{14}$ and $10^{15} \,h^{-1} M_{\rm 500}$, respectively. We note that the values for the temperature of the intercluster medium obtained in this way are in agreement with numerical simulations ($10^5$--$10^7$ K; see e.g. Cen \& Ostriker 1999; Dav{\'e} et al. 2001) and observations (e.g. Williams et al. 2007) of the warm-hot intergalactic medium.

Alternatively, we can relax the assumption of isolation and consider the environment of groups and clusters, effectively considering different values of $T_{\rm \infty}$ for each system. Whereas local clusters, given their size, can almost always be considered as isolated systems, most galaxy groups, on the other hand, are believed to exist embedded in proto-clusters (or even in clusters). Therefore, the effective $T_{\rm \infty}$ of groups tends to be higher than that of clusters, causing a steepening of the $M$--$T$ relation. Fossil groups, on the other hand, can be considered to be fairly isolated, hence to have a lower effective $T_{\rm \infty}$ than non-isolated groups (cf. Khosroshahi, Ponman \& Jones 2007). This range in the environment of low-mass systems -- from isolated, to embedded in proto-clusters, to embedded in clusters -- could be one of the possible causes for the larger scatter of the $M$--$T$ relation at low temperatures than at high temperatures (see Section \ref{sec:Dependence_on_the_mass-concentration_relation} for another possible cause). Likely, these two interpretations (mass-dependent and environment-dependent) co-exist.

Finally, if we allow for the truncation of the DM gravitational potential or for the breaking of the HE assumption at a given, finite radius (e.g. $\xi\,r_{\rm vir}$), the polytropic solution is not global any more. In this case, $T_{\rm \xi}$ can be interpreted as being related to the external surface pressure term in the virial theorem, as Ostriker et al. (2005) assumed in the case $\xi =1$.

%SECTION 3.1.3
%SECTION 3.1.3
%SECTION 3.1.3
%SECTION 3.1.3
%SECTION 3.1.3

\subsubsection{Dependence on the DM logarithmic slope at large radii}\label{sec:Dependence_on_the_DM_logarithmic_slope_at_large_radii}

\begin{figure}
\centering
\vspace{5pt}
\includegraphics[width=0.72\columnwidth,angle=90]{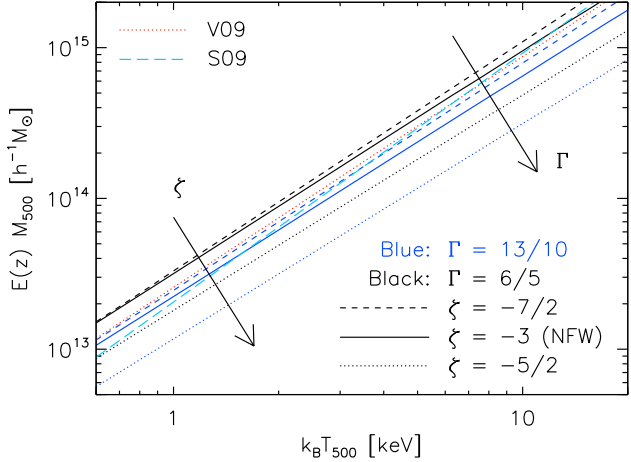}
\vspace{5pt}
\caption{Constraints on the DM logarithmic slope at large radii, for systems at redshift $z=0$ ($E(z)=1$), with B10 DM, concentration given by D08 and a polytropic gas with $T_{\rm \infty}=0$, via comparison between our theoretical model and observations by V09 and S09 (\emph{red, dotted} and \emph{turquoise, long-dashed lines}, respectively). We plot the theoretical $M$--$T$ relation for two polytropic models -- $\Gamma=6/5$ (\emph{black lines}) and $\Gamma=13/10$ (\emph{blue lines}) -- and three DM models -- $\zeta=-5/2$ (\emph{dotted lines}), $\zeta=-3$ (NFW; \emph{solid lines}) and $\zeta=-7/2$ (\emph{dashed lines}). Assuming a mass-independent polytropic index $6/5 \lesssim \Gamma \lesssim 13/10$ for high-mass systems, we obtain $-3.5 \lesssim \zeta \lesssim -2.7$: the DM in galaxy clusters either follows the NFW profile or is slightly steeper at large radii.}
\label{fig:MT_Relation_with_beta_from_observations}
\end{figure}

In this section, we consider systems at redshift $z=0$ ($E(z)=1$), described by a generalised NFW DM profile (B10), with concentration given by D08 and a polytropic gas with a given $\Gamma$ and $T_{\rm \infty}=0$, and study the dependence of the $M$--$T$ relation on the DM logarithmic slope at large radii $\zeta$.

To achieve this, we study the effects of varying $\beta \equiv -\zeta-1$ in Equation \eqref{eq:DM_rho}. The B10 profile is itself a simplified version of the general profile given in Suto et al. (1998). However, Suto et al. (1998) themselves, and several authors thereafter (including KS01 and KS02), use a simplified version of their own profile (see also Zhao 1996), by fixing the large-radii logarithmic slope to $\zeta=-3$. The advantage of using the B10 profile is that we can compare our model to observations and obtain constraints on how DM behaves at large radii. See Appendix \ref{sec:Overdensities} for the computation of different overdensity masses and radii in the case of a B10 profile.

Comparing the results of our model to recent X-ray observations of clusters (V09), we can discriminate between different values of $\zeta$, assumed to be the same for systems at all masses. In Figure \ref{fig:MT_Relation_with_beta_from_observations}, we show the $M$--$T$ relation for two values of the polytropic index and for three values of $\zeta$. Assuming that a mass-independent polytropic index has a value $6/5 \lesssim \Gamma \lesssim 13/10$, we obtain $-3.5 \lesssim \zeta \lesssim -2.7$: the DM in galaxy clusters either follows the NFW profile or is slightly steeper at large radii. The comparison to low-mass systems (S09) is less straightforward: a constant value of $\zeta$ is not consistent with the data. However, more causes could be at work, including a mass-dependent polytropic index and/or a non-zero temperature at large radii (see Sections \ref{sec:Dependence_on_the_polytropic_index} and \ref{sec:Dependence_on_the_gas_temperature_at_large_radii}), hence complicating the determination of $\zeta$.

As expected, for a fixed mass, the temperature is higher for lower values of $\beta$ (less negative values of $\zeta$). This is easily understood by investigating Equation \eqref{eq:T_0_thr} or the definition of $\phi_{\rm 0}$, which is proportional to $1/(\beta -1)$: the lower $\beta$ is, the deeper the central gravitational potential and therefore the higher the central temperature has to be, in order for the gas to be unbound or bound only in a limitless region. The relation between the central temperature and the average temperature is not, however, a linear relation. The parameter $\beta$ also affects the shape of the temperature profile (through the gas parameter $\Delta_{\rm gas}$, given by Equation \ref{eq:gas_parameter}). For this reason, the $M$--$T$ relation does not shift indefinitely towards \emph{lower} values of temperature (for a given mass). Instead, for values of $\beta \gtrsim 5/2$, it slightly shifts towards \emph{higher} values of temperature, especially at lower masses (not shown in the figure).

One can speculate on the interdependence between $\zeta$ and $\Gamma$, hinted at by Figure \ref{fig:MT_Relation_with_beta_from_observations}. Does the gas behave differently depending on the gravitational potential -- i.e. do we have a $\Gamma(\zeta)$ relation? -- or does the DM respond somehow to the gas -- i.e. $\zeta(\Gamma)$? Even though it is possible to think of scenarios where DM is significantly affected by baryons (e.g. adiabatic contraction being one of them), it is unlikely that such physics is this effective at large scales. If there is indeed a direct relation between the polytropic index of the gas and the logarithmic slope of DM at large radii, it is more likely that the gas is affected by DM: $\Gamma(\zeta)$. This is consistent with the explanation given by Bertschinger (1985) for the existence of a polytropic relation in the first place: gas adjusts itself according to the formation process of the DM halo, and the formation process has an effect on the value of $\zeta$ (e.g. Lu et al. 2006).

%SECTION 3.1.4
%SECTION 3.1.4
%SECTION 3.1.4
%SECTION 3.1.4
%SECTION 3.1.4

\subsubsection{Dependence on the mass-concentration relation}\label{sec:Dependence_on_the_mass-concentration_relation}

\begin{figure}
\centering
\vspace{5pt}
\includegraphics[width=0.99\columnwidth,angle=0]{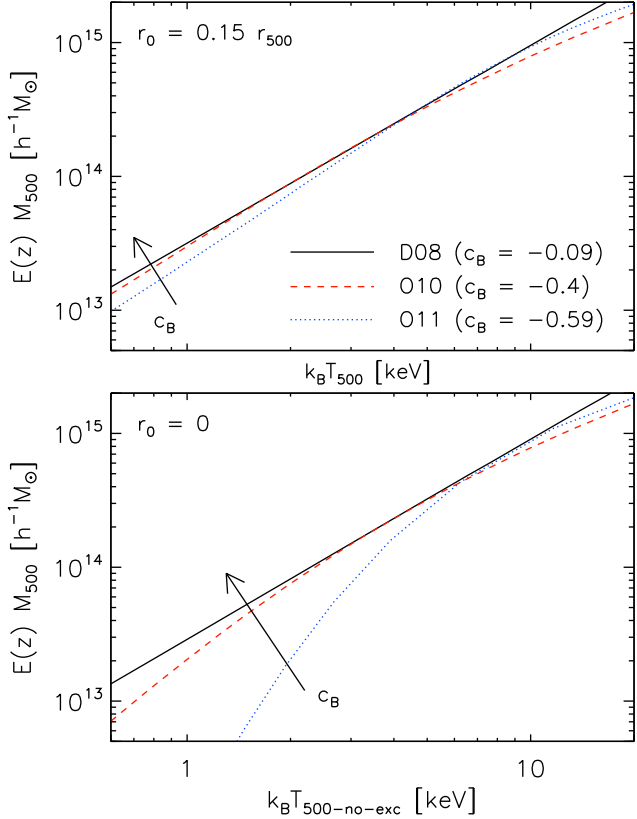}
\vspace{-4pt}
\caption{Dependence of the theoretical $M$--$T$ relation on the $c_{\rm vir}$--$M_{\rm vir}$ relation, for systems at redshift $z=0$ ($E(z)=1$), with polytropic gas in HE with NFW DM ($\zeta=-3$) at all radii, $\Gamma=6/5$ and $T_{\rm \infty}=0$. \emph{Black, solid line}: concentration given by D08. \emph{Red, dashed line}: concentration given by O10. \emph{Blue, dotted line}: concentration given by O11. {\bf Top panel}: excision ($r_{\rm 0}=0.15\,r_{\rm 500}$ in the computation of $T_{\rm 500}$). Different mass-concentration relations do not cause significant differences in the $M$--$T$ relation. {\bf Bottom panel}: no excision ($r_{\rm 0}=0$). The effect is much stronger: excision clearly discards information.}
\label{fig:MT_Relation_with_cvir}
\end{figure}

In this section, we consider systems at redshift $z=0$ ($E(z)=0$), described by an NFW DM profile ($\zeta=-3$), and a polytropic gas with $\Gamma=6/5$ and $T_{\rm \infty}=0$, and study the dependence of the $M$--$T$ relation on the concentration $c_{\rm vir}$.

Numerical simulations and observations show that concentration is a function of mass and redshift (see e.g. Equation \ref{eq:c_vir_M_vir_relation}). We focus here on the dependence of concentration on mass. Given the vast amount of published results and the discordance amongst them, we chose three representative mass-concentration relations with three fairly different dependences on mass: $c_{\rm B}=-0.09$ (D08), $c_{\rm B}=-0.4$ (Okabe et al. 2010: O10) and $c_{\rm B}=-0.59$ (Oguri et al. 2011: O11). Amongst these relations, there is substantial agreement in the value of the concentration of high-mass systems. Concentrations for low-mass systems, however, differ significantly\footnote{As in the previous sections, we assume Equation \eqref{eq:c_vir_M_vir_relation} to be valid for all virial masses between $10^{12}$M$_{\rm \odot}$ and $10^{16}$M$_{\rm \odot}$. D08 used redshift-dependent DM-only simulations, whereas O10 and O11 used lensing observations of clusters at mean redshift $z_{\rm mean}=0.23$ and 0.45, respectively. Assuming the same redshift dependence for all three relations (i.e. $c_{\rm C}=-0.69$), we then normalised the results of O10 and O11 by dividing their parameter $c_{\rm A}$ by $f_{\rm DM}^{c_{\rm B}}(1+z_{\rm mean})^{c_{\rm C}}$.}.

In Figure \ref{fig:MT_Relation_with_cvir}, we show the $M$--$T$ relation for the three mentioned mass-concentration relations. Surprisingly, very different concentrations do not translate into significant differences in the $M_{\rm 500}$--$T_{\rm 500}$ relation, except in the very high-temperature regime, which is difficult to analyse, since very few clusters live there. Different concentrations imply varied gravitational potentials, and one would expect the gas to respond accordingly, giving rise to sizeably different $M$--$T$ relations. This is a clear example where excision discards important information. If we compute the $M_{\rm 500}$--$T_{\rm 500-no-exc}$ relation instead (by replacing $r_{\rm 0}=0.15\,r_{\rm 500}$ with $r_{\rm 0}=0$ in Equation \ref{eq:T_average}), we obtain a much stronger effect: at low masses, the temperature is higher for lower (more negative) $c_{\rm B}$. The effect of concentration is sizeable only in the inner-most region of the system. By performing excision to remove scatter, one also removes any information on the concentration. Moreover, part of the scatter of the $M$--$T$ relation at low temperatures might be due to the intrinsic scatter in the mass-concentration relation. We additionally note that excision also removes information on different DM inner slopes (e.g. Moore et al. 1998) or on the effect of stars on the gravitational potential, especially at low masses (e.g. Capelo et al. 2010).

Incidentally, we notice that the mass-concentration relation might be partially responsible for the steepening of the $M$--$T$ relation (with excision) at low masses. However, as we will see in Section \ref{sec:The_L_T_Relation}, mass-concentration relations with a strong mass dependence (like O10 and O11) are ruled out by the comparison of the model to $L\,$--$\,T$ observations.

%SECTION 3.1.5
%SECTION 3.1.5
%SECTION 3.1.5
%SECTION 3.1.5
%SECTION 3.1.5

\subsubsection{Dependence on redshift}\label{sec:Dependence_on_redshift}

\begin{figure}
\centering
\vspace{5pt}
\includegraphics[width=0.72\columnwidth,angle=90]{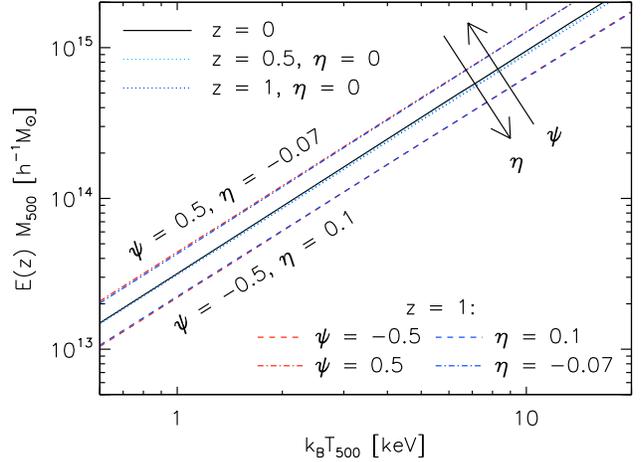}
\vspace{5pt}
\caption{Dependence of the theoretical $M$--$T$ relation on redshift, for systems with polytropic gas in HE with NFW DM ($\zeta=-3$) at all radii, concentration given by D08, $\Gamma=6/5$ and $T_{\rm \infty}=0$. We plot the $M$--$T$ relation for redshifts $z=0$ (\emph{black, solid line}), 0.5 (\emph{turquoise, dotted line}) and 1 (\emph{blue, dotted line}). The overlap of these three lines shows that the scaling of the model with redshift is very close to being self-similar. We also plot the $M$--$T$ relation at redshift $z=1$ for two different cases of non-self-similarity (see text for more details): $\psi=-0.5$ (\emph{red, dashed line}) and $0.5$ (\emph{red, dot-dashed line}). These two relations are well fit by our model when $\Gamma$ varies with redshift according to Equation \eqref{eq:Gamma_of_z_Clerc}, when $\eta$ is given by Equation \eqref{eq:eta_of_psi} (\emph{blue lines}). The same (not shown in the figure) can be achieved using Equations \eqref{eq:Gamma_of_z_Reichert} and \eqref{eq:theta_of_upsilon}.}
\label{fig:MT_Relation_with_z}
\end{figure}

In this section, we consider systems described by an NFW DM profile ($\zeta = -3$), with concentration given by D08 and a polytropic gas with $\Gamma=6/5$ and $T_{\rm \infty}=0$, and study the dependence of the $M$--$T$ relation on redshift.

If we assume all galaxy group and cluster potentials to have the same shape (i.e. a fixed concentration) then, from the virial theorem, $T_{\rm \Delta_{\rm c}} \propto M_{\rm \Delta_{\rm c}}/r_{\rm \Delta_{\rm c}} \propto M_{\rm \Delta_{\rm c}}^{2/3} \rho_{\rm mean}^{1/3}$, where $\rho_{\rm mean}=\Delta_{\rm c}\rho_{\rm c}$ is the mean density within $r_{\rm \Delta_{\rm c}}$. Since $\rho_{\rm c} \propto E^2(z)$, we expect

\begin{equation}\label{eq:T_of_z}
T_{\rm \Delta_{\rm c}} \propto
    \begin{cases}\displaystyle{M_{\rm \Delta_{\rm c}}^{2/3}E^{2/3}(z)} & \mbox{if $\Delta_{\rm c}=\hbox{fixed number}$,} \cr
                 \displaystyle{M_{\rm \Delta_{\rm c}}^{2/3}\Delta_{\rm vir}^{1/3}(z)E^{2/3}(z)} & \mbox{if $\Delta_{\rm c} \propto \Delta_{\rm vir}(z)$.} \cr
    \end{cases}
\end{equation}

If we assume a flat, matter-dominated Universe (i.e. $\Omega_{\rm 0}=1$), then $E(z) \propto (1+z)^{3/2}$ and we recover the results by Kaiser (1986), where $T_{\rm \Delta_{\rm c}} \propto M_{\rm \Delta_{\rm c}}^{2/3}(1+z)$.

In Figure \ref{fig:MT_Relation_with_z}, we compare the results of our model at high redshift ($z=0.5$ and $z=1$) to those at $z=0$, finding that the scaling of the model with redshift is very close to self-similar, according to Equation \eqref{eq:T_of_z}, even when using a concentration that varies with mass and redshift. Qualitatively, the effect of varying redshift is consistent with what already shown in Figure \ref{fig:Dependence_on_z_varying_T}: for a fixed $\Gamma$, when $z$ increases, the threshold central temperature increases, and so does the average temperature. We do not extend the model to higher redshift, because the assumptions of HE and isolation become less and less valid with increasing $z$, due to the fact that clusters are actively assembling at these epochs.

In the past decade, several high-redshift X-ray observations of clusters have been conducted (e.g. Reichert et al. 2011 and references therein). In some cases, the self-similarity of the $M$--$T$ relation has been observed (e.g. Ettori et al. 2004; Reichert et al. 2011), and the theoretical model with the parameters listed above is in good agreement with it.

However, there exist observations (e.g. Jee et al. 2011; Clerc et al. 2011; see also numerical simulations by Short et al. 2010) which show a discrepancy in the normalisation of the $M$--$T$ relation between the observed values and the values expected from self-similarity.

If we assume that our model is accurate also at high redshift, there are two plausible ways to recover the observed non-self-similar relations. Assuming that the polytropic index does not vary with time, one needs the DM profile to depend on redshift (e.g. $\zeta(z)$). Assuming instead that the DM profile does not evolve, one needs some variation of the polytropic index with redshift (e.g. $\Gamma(z)$). These two causes might be concurrent. Here, we focus on the latter possibility and present a simple method to obtain the $\Gamma\,$--$\,z$ relation from observations.

It is customary to parametrize the observed discrepancy from self-similarity by adding a redshift-dependent term to the $M$--$T$ relation:

\begin{equation}\label{eq:non_self_similar_M_T}
E(z)M_{\rm \Delta_{\rm c}} \propto T_{\rm \Delta_{\rm c}}^{\alpha} f_{\rm M-T}(z),
\end{equation}

\noindent where the value of $\alpha$, assumed to be the same at all redshifts, does not affect the results of this section, and $f_{\rm M-T}(z)$ varies according to the author (with $f_{\rm M-T}(z=0)=1$).

We now parametrize the time-dependence of the polytropic index accordingly and write

\begin{equation}\label{eq:Gamma_of_z}
\Gamma(z)=\Gamma_0 f_{\rm \Gamma}(z),
\end{equation}

\noindent where $\Gamma_0 \equiv \Gamma(z=0)$ is the local value of the polytropic index and $f_{\rm \Gamma}(z=0)=1$.

By combining together Equations \eqref{eq:MT_relation_Gamma}, \eqref{eq:non_self_similar_M_T} and \eqref{eq:Gamma_of_z}, our theoretical model reproduces the discrepancy from self-similarity when

\begin{equation}\label{eq:f_Gamma_of_z}
f_{\rm \Gamma}(z) = \frac{1}{\Gamma_0}\frac{\Gamma_0f_{\rm M-T}(z)^{1/\epsilon}/(\Gamma_0-1)}{[\Gamma_0f_{\rm M-T}(z)^{1/\epsilon}/(\Gamma_0-1)]-1}.
\end{equation}

The most common parametrizations for the non-self-similar $M$--$T$ relation are (Case A; e.g. Short et al. 2010; Clerc et al. 2011)

\begin{equation}\label{eq:Clerc2011}
E(z)M_{\rm \Delta_{\rm c}} \propto T_{\rm \Delta_{\rm c}}^{\alpha} (1+z)^{\psi}
\end{equation}

\noindent and (Case B; e.g. Reichert et al. 2011)

\begin{equation}\label{eq:Reichert2011}
E(z)^{\upsilon}M_{\rm \Delta_{\rm c}} \propto T_{\rm \Delta_{\rm c}}^{\alpha},
\end{equation}

\noindent where $\psi$ and $\upsilon$ quantify the discrepancy from the self-similar case\footnote{These two parametrizations give very similar results only in the range $z=0\,$--$\,1$, which is exactly the range we consider in this section. At $z=1$, Equation \eqref{eq:Clerc2011} with $\psi=0.5$ or $-0.5$ is similar to Equation \eqref{eq:Reichert2011} with $\upsilon=0.33$ or $1.67$, respectively.}. For self-similarity, one needs $\psi=0$ and $\upsilon=1$. The above descriptions imply $f_{\rm M-T}(z)=(1+z)^{\psi}$ and $f_{\rm M-T}(z)=E(z)^{1-\upsilon}$, respectively, in Equation \eqref{eq:non_self_similar_M_T}. We now parametrize the time-dependence of the polytropic index accordingly. We assume $f_{\rm \Gamma}(z)=(1+z)^{\eta}$ when using $f_{\rm M-T}(z)=(1+z)^{\psi}$, and $f_{\rm \Gamma}(z)=E(z)^{\theta}$ when using $f_{\rm M-T}(z)=E(z)^{1-\upsilon}$, where $\eta$ and $\theta$ quantify the evolution of $\Gamma$. For self-similar models, one has $\eta=\theta=0$.

The resulting $\Gamma\,$--$\,z$ relation is then (Case A)

\begin{equation}\label{eq:Gamma_of_z_Clerc}
\Gamma(z)=\Gamma_0(1+z)^{\eta},
\end{equation}

\noindent where

\begin{equation}
\eta \log_{\rm 10}(1+z) = \log_{\rm 10}\left[\frac{(1+z)^{\psi/\epsilon}/(\Gamma_0-1)}{[\Gamma_0(1+z)^{\psi/\epsilon}/(\Gamma_0-1)]-1}\right],
\end{equation}

\noindent or (Case B)

\begin{equation}\label{eq:Gamma_of_z_Reichert}
\Gamma(z)=\Gamma_0E(z)^{\theta},
\end{equation}

\noindent where

\begin{equation}
\theta \log_{\rm 10}E(z) = \log_{\rm 10}\left[\frac{E(z)^{(1-\upsilon)/\epsilon}/(\Gamma_0-1)}{[\Gamma_0E(z)^{(1-\upsilon)/\epsilon}/(\Gamma_0-1)]-1}\right].
\end{equation}

These relations are very weakly dependent on redshift, especially in the redshift range $z=0\,$--$\,1$ and for low values of $\psi$. Hence, after fixing the other parameters (e.g. $\epsilon=\Gamma_0=6/5$, see Section \ref{sec:Dependence_on_the_polytropic_index}), we can provide a relation between $\eta$ and $\psi$,

\begin{equation}\label{eq:eta_of_psi}
\eta = -0.17\,\psi+0.06\,\psi^2,
\end{equation}

\noindent and between $\theta$ and $\upsilon$,

\begin{equation}\label{eq:theta_of_upsilon}
\theta = -0.126 + 0.082 \,\upsilon + 0.044 \,\upsilon^2.
\end{equation}

Observations of non-self-similarity are still giving very disparate results (e.g. Reichert et al. 2011; Clerc et al. 2011). In the future, when better observations will be at hand and the value of $\psi$ will be known to higher accuracy than now, it will be possible to infer the rate of evolution of the polytropic index, by using e.g. Equations \eqref{eq:Gamma_of_z_Clerc} and \eqref{eq:eta_of_psi}. This will also let us compare our results to numerical simulation (e.g. Battaglia et al. 2011b).

%SECTION 3.2
%SECTION 3.2
%SECTION 3.2
%SECTION 3.2
%SECTION 3.2

\subsection{The $L\,$--$\,T$ relation}\label{sec:The_L_T_Relation}

\begin{figure}
\centering
\vspace{5pt}
\includegraphics[width=0.99\columnwidth,angle=0]{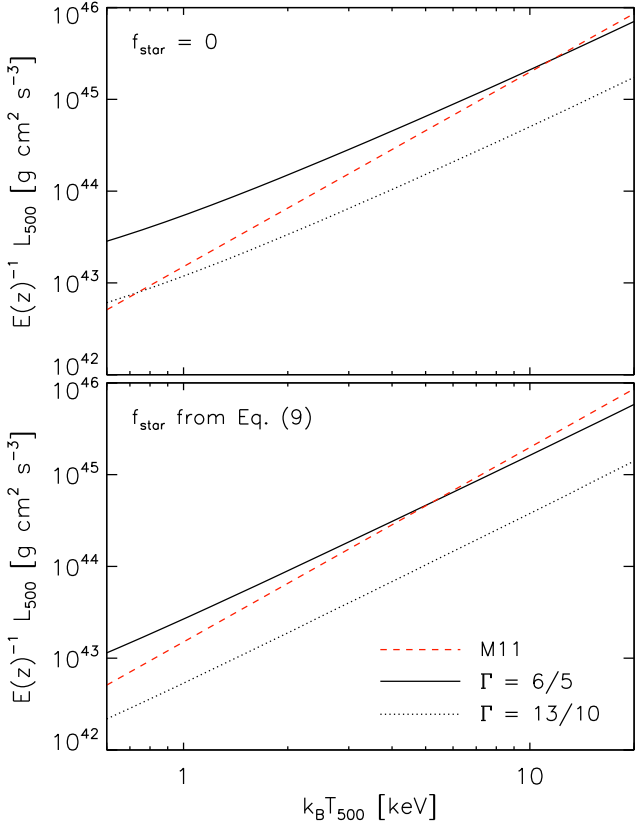}
\vspace{-4pt}
\caption{Dependence of the theoretical $L\,$--$\,T$ relation on the polytropic index and on the gas fraction, for systems at redshift $z=0$ ($E(z)=1$), with polytropic gas in HE with NFW DM ($\zeta=-3$) at all radii, concentration given by D08 and $T_{\rm \infty}=0$. \emph{Black, solid lines}: $\Gamma=6/5$. \emph{Black, dotted lines}: $\Gamma=13/10$. \emph{Red, dashed lines}: observations by M11. {\bf Top panel}: $f_{\rm gas}=f_{\rm b}$ ($f_{\rm star}=0$). {\bf Bottom panel}: $f_{\rm gas}$ varies ($f_{\rm star}$ is given by Equation \ref{eq:f_star}).}
\label{fig:LT_Relation_D08}
\end{figure}

\begin{figure}
\centering
\vspace{5pt}
\includegraphics[width=0.99\columnwidth,angle=0]{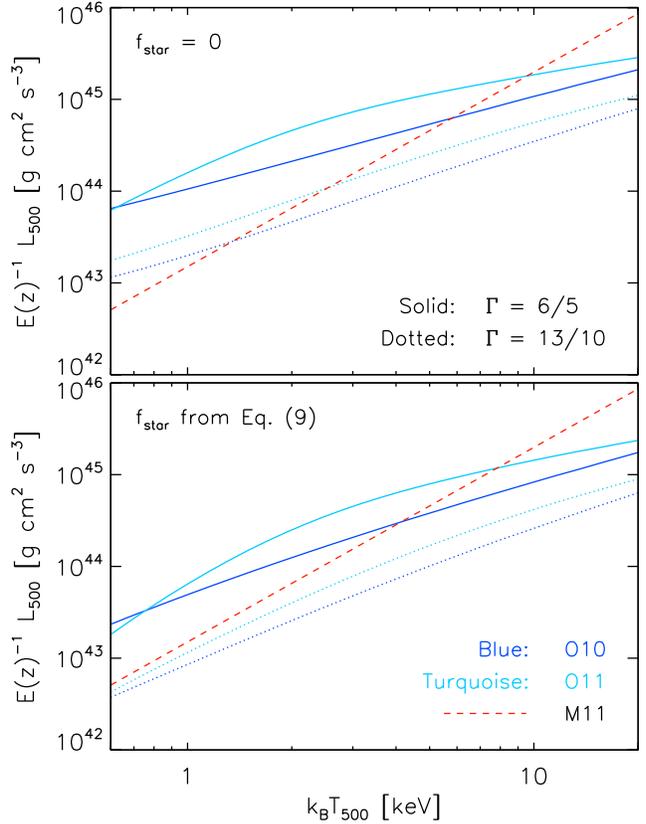}
\vspace{-4pt}
\caption{Dependence of the theoretical $L\,$--$\,T$ relation on the polytropic index, the gas fraction and the concentration, for systems at redshift $z=0$ ($E(z)=1$), with polytropic gas in HE with NFW DM ($\zeta=-3$) at all radii and $T_{\rm \infty}=0$. \emph{Solid lines}: $\Gamma=6/5$. \emph{Dotted lines}: $\Gamma=13/10$. \emph{Blue lines}: concentration given by O10. \emph{Turquoise lines}: concentration given by O11. \emph{Dashed, red lines}: observations by M11. {\bf Top panel}: $f_{\rm gas}=f_{\rm b}$ ($f_{\rm star}=0$). {\bf Bottom panel}: $f_{\rm gas}$ varies ($f_{\rm star}$ is given by Equation \ref{eq:f_star}).}
\label{fig:LT_Relation_O10_O11}
\end{figure}

In this section, we derive the $L\,$--$\,T$ relation from our model and compare it to the observed relation of a sample of relaxed clusters: $E(z)^{-1} L_{\rm 500} = (6.71 \pm 0.34) \times 10^{44}$ g cm$^2$ s$^{-3}$ ($k_{\rm B}T_{\rm 500}/6$ keV)$^{2.12 \pm 0.17}$ (M11). To be consistent with how we defined the temperature (in Section \ref{sec:Average_temperature_and_the_issue_of_excision}), we set

\begin{equation}\label{eq:Luminosity}
L_{\rm 500} = \int_{\rm 0.15\,r_{\rm 500}}^{r_{\rm 500}}n_{\rm e}n_{\rm i}\Lambda_{\rm N}4\pi r^2dr,
\end{equation}

\noindent where the cooling function $\Lambda_{\rm N}$ is described in Appendix \ref{sec:Cooling_Function}.

In Section \ref{sec:The_M_T_Relation}, we determined the temperature of groups and clusters of galaxies by imposing HE at all radii, between a polytropic gas and a given DM gravitational potential. For that calculation, only two constraints were necessary, namely on the polytropic index and on the central gas temperature, because we did not include gas self-gravity in the calculations. In this section, when deriving the luminosity, a third constraint (on the central gas density) is necessary. Different authors use different methods: KS01 (and KS02) impose the gas density at large radii to be $f_{\rm b}$ times the density of DM; Ascasibar et al. (2003) require the baryon fraction to never exceed the cosmic value; A06 assume the baryon fraction at three DM scale radii to be equal to $\Omega_{\rm b}/\Omega_{\rm 0}$; Ostriker et al. (2005; see also Bode et al. 2007, 2009) impose conservation of energy.

In this paper, the constraint on $\rho_{\rm 0}$ is given by imposing the baryon fraction at the virial radius to be equal to the cosmic value, independent of mass or redshift. Therefore, the model depends also on the stellar content, and we compare results assuming the stellar mass fraction is given either by Equation \eqref{eq:f_star} or by $f_{\rm star}=0$ (i.e. $f_{\rm gas}=f_{\rm b}$).

In Figure \ref{fig:LT_Relation_D08}, we show the theoretical $L\,$--$\,T$ relation for two values of the polytropic index and one mass-concentration relation (D08), and compare it to recent observations by M11, for both cases of constant and mass-dependent gas mass fraction. We note how the $L\,$--$\,T$ relation is well approximated by a power law only at high temperatures, when assuming a constant gas mass fraction, and is well described by a power law at all temperatures, when assuming a mass-dependent gas mass fraction. The effect of a gas mass fraction dependent on the total mass of the system is clearly visible, especially at low masses. This is a potential source of scatter, since different systems having the same total (low) virial mass might have slightly different stellar masses.

When assuming a mass-dependent gas mass fraction, the $L\,$--$\,T$ relation with $\Gamma=6/5$ matches X-ray observations better than the case with $\Gamma=13/10$.

Focusing on other mass-concentration relations (e.g. O10 or O11, see Figure \ref{fig:LT_Relation_O10_O11}), we immediately notice that the model does not recover the observed $L\,$--$\,T$ relation by M11. This might imply that concentration cannot depend strongly on mass.

%SECTION 4
%SECTION 4
%SECTION 4
%SECTION 4
%SECTION 4

\section{Summary and conclusions}\label{sec:Conclusions}

We present a simple and flexible model for the gas and DM content of groups and clusters of galaxies, by assuming an ideal polytropic gas in HE at all radii with a DM gravitational potential, described by a generalised NFW profile. Exploring the physical properties of the polytropic solution of the HE equation, we report on methods to constrain the three gas parameters ($\Gamma$, $T_{\rm 0}$ and $\rho_{\rm 0}$) and two DM parameters ($\zeta$ and $c_{\rm vir}$) of such a solution. With no prior knowledge of the polytropic index or the boundary conditions, we derive the scaling relations for galaxy groups and clusters from these first principles and compare them to recent X-ray observations. This allows us to obtain powerful constraints on gas and DM parameters. In particular,

\begin{itemize}
\item for systems at redshift $z=0$, with NFW DM, concentration given by D08 and $T_{\rm \infty}=0$, we derive a theoretical $M$--$T$ relation with a logarithmic slope of 1.46 and normalisation that depends rather simply on the polytropic index: $[\Gamma/(\Gamma-1)]^{6/5}$ (see Equation \ref{eq:MT_relation_Gamma});
\item comparing with recent X-ray observations of clusters of galaxies, we find that a model with a constant polytropic index with a value between 6/5 and 13/10 is consistent with the data;
\item further, when including data from low-mass systems, the assumption of a constant polytropic index is not consistent with the observations. A possible explanation for the discrepancy is a mass dependence in $\Gamma$ such that it decreases with mass (and increases with concentration), as derived in Equations \eqref{eq:Gamma_M_relation} and \eqref{eq:Gamma_c_relation};
\item for systems at redshift $z=0$, with NFW DM, concentration given by D08 and $\Gamma=6/5$, we derive a theoretical $M$--$T$ relation that is a broken power law, where the break depends on the value of the temperature at large radii. Comparing this case to X-ray observations of groups, we constrain the temperature at large radii to be $10^6$ K $\lesssim T_{\rm \infty} \lesssim 10^7$ K, consistent with numerical simulations and observations of the warm-hot intergalactic medium;
\item for systems at redshift $z=0$, with a generalised NFW DM, concentration given by D08 and $T_{\rm \infty}=0$, we compare the theoretically derived $M$--$T$ relation to recent X-ray observations of clusters of galaxies. Assuming a mass-independent polytropic index $6/5 \lesssim \Gamma \lesssim 13/10$ for high-mass systems, we find that $-3.5 \lesssim \zeta \lesssim -2.7$, i.e. the DM profile in galaxy clusters either follows the NFW profile or is slightly steeper at large radii;
\item for systems at redshift $z=0$, with NFW DM, $\Gamma=6/5$ and $T_{\rm \infty}=0$, we also study the dependence of the theoretically derived $M$--$T$ relation on concentration and find a very weak dependence, unless the temperature is calculated without excising the central region;
\item for systems with NFW DM, concentration given by D08, $\Gamma=6/5$ and $T_{\rm \infty}=0$, we study the dependence of the theoretically derived $M$--$T$ relation on redshift, and find that our model scales in a self-similar behaviour. However, given the freedom in our model, the departure from self-similarity seen in some clusters can be used to constrain the evolution of the polytropic index with redshift (see e.g. Equations \ref{eq:Gamma_of_z_Clerc} and \ref{eq:eta_of_psi});
\item for systems at redshift $z=0$, with NFW DM and $T_{\rm \infty}=0$, we study the dependence of the theoretically derived $L\,$--$\,T$ relation on the polytropic index, on concentration and on the stellar mass fraction. We find that a concentration strongly dependent on mass is ruled out by comparison to recent X-ray observations.
\end{itemize}

The power of this model lies in its simplicity and its flexibility. In future work, more sophisticated models can be built upon this framework by relaxing some of the assumption made here, e.g. spherical symmetry, and by including non-thermal contributions to HE; a radially-dependent polytropic index; more general DM profiles; gas self-gravity; and the effect of the stellar gravitational potential. We also note that we studied narrow slices of the five-dimensional parameter space ($\Gamma$, $T_{\rm 0}$, $\rho_{\rm 0}$, $\zeta$ and $c_{\rm vir}$). Further studies can easily exploit the model to explore the full parameter space.

Although we have used the polytropic relation to describe the gas, its origin is still unclear, although it is likely tied to the formation of the halo. In this paper, we have proposed a relation between the polytropic index, mass and redshift, thus providing useful information and a first step to solve this puzzle.

\section*{Acknowledgments}

PRC is grateful for helpful discussions with Daisuke Nagai and Frank C. van den Bosch. PN acknowledges support from the National Science Foundation under grant AST10-44455.

%\newpage
%\hspace{1cm}
%\newpage

\appendix

\section{Overdensity masses and radii}\label{sec:Overdensities}

\begin{figure}
\centering
\vspace{5pt}
\includegraphics[width=0.99\columnwidth,angle=0]{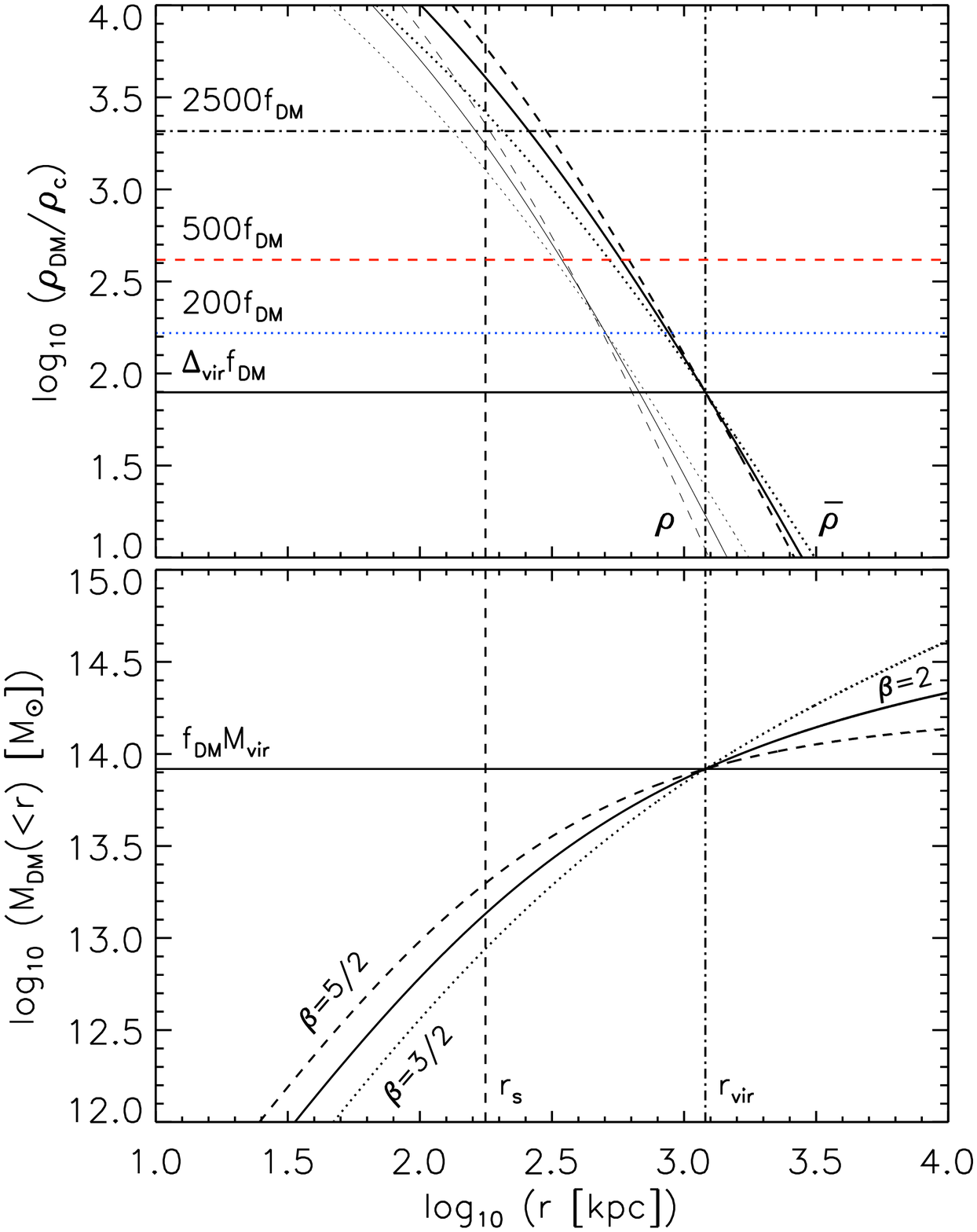}
\vspace{-4pt}
\caption{DM radial profiles for a system of virial mass $M_{\rm vir}=10^{14}$M$_{\rm \odot}$ at redshift $z=0$, with concentration given by D08, for three cases of the B10 profile: $\beta=3/2$ (\emph{black, dotted curves}), $\beta=2$ (the NFW case; \emph{black, solid curves}) and $\beta=5/2$ (\emph{black, dashed curves}). The \emph{black, vertical, dot-dashed lines} denote the virial radius, whereas the \emph{black, vertical, dashed lines} denote the DM scale radius. {\bf Top panel}: DM density (\emph{thin curves}) and average density (\emph{thick curves}) radial profiles. The \emph{horizontal lines} denote the $\Delta_{\rm vir}f_{\rm DM}\rho_{\rm c}$ (\emph{black, solid}), $200f_{\rm DM}\rho_{\rm c}$ (\emph{blue, dotted}), $500f_{\rm DM}\rho_{\rm c}$ (\emph{red, dashed}) and $2500f_{\rm DM}\rho_{\rm c}$ (\emph{black, dot-dashed}) density values, respectively. {\bf Bottom panel}: DM enclosed mass radial profiles. The \emph{black, horizontal, solid line} denotes the DM virial mass.}
\label{fig:NFW_and_Bulbul_profiles}
\end{figure}

\begin{figure}
\centering
\vspace{5pt}
\includegraphics[width=0.99\columnwidth,angle=0]{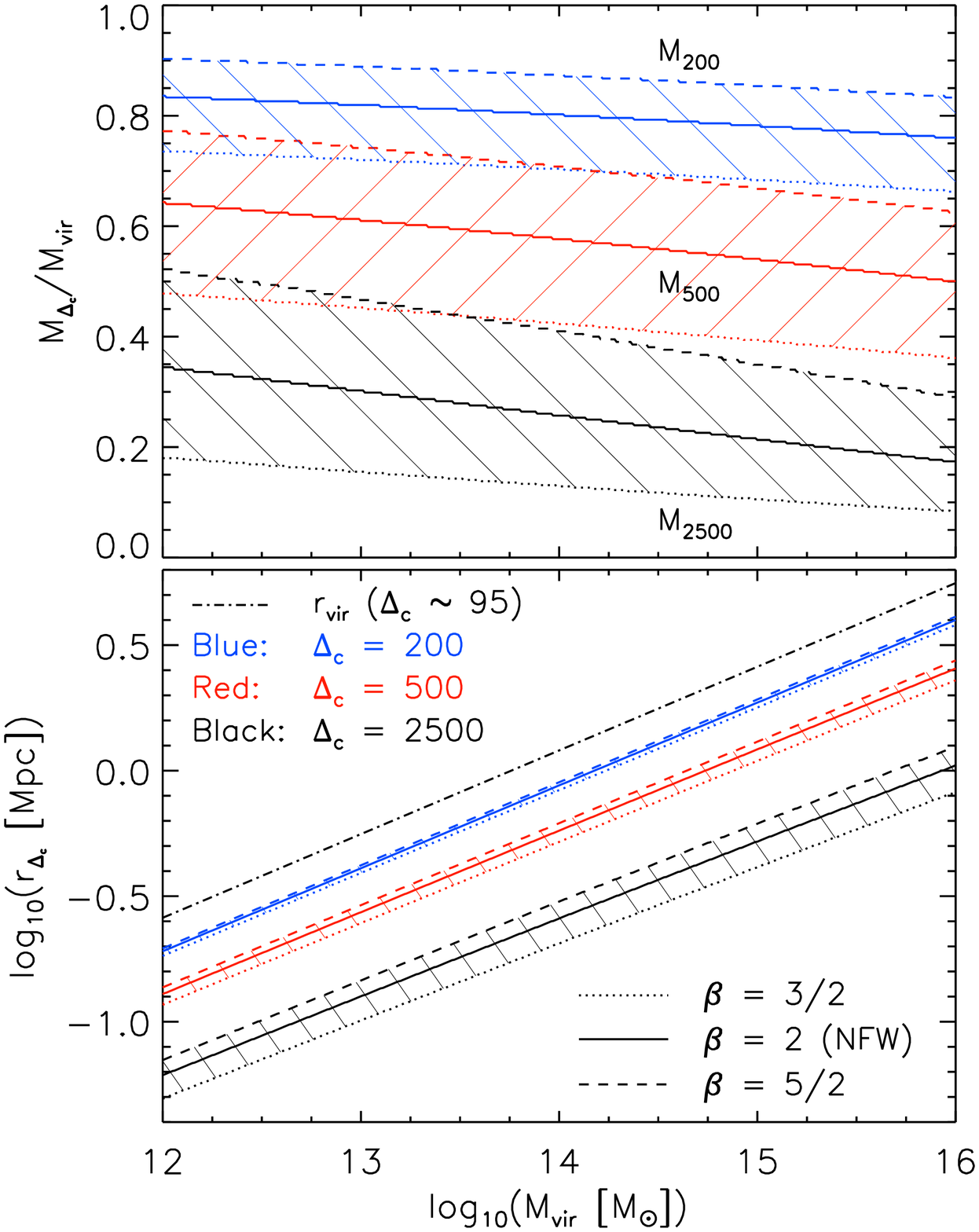}
\vspace{-4pt}
\caption{DM overdensity masses (top panel) and radii (bottom panel), for systems at redshift $z=0$, with concentration given by D08, for three cases of the B10 profile ($\beta=3/2$ -- \emph{dotted lines}; $\beta=2$ -- the NFW case, \emph{solid lines}; and $\beta=5/2$ -- \emph{dashed lines}) and three values of $\Delta_{\rm c}$ ($\Delta_{\rm c}=200$ -- \emph{blue lines}; $\Delta_{\rm c}=500$ -- \emph{red lines}; and $\Delta_{\rm c}=2500$ -- \emph{black lines}). The \emph{black, dot-dashed line} in the bottom panel is the virial radius.}
\label{fig:Overdensities}
\end{figure}

In this section, we describe the difference between an NFW profile and a B10 profile with $\beta \neq 2$. In Figure \ref{fig:NFW_and_Bulbul_profiles}, we show the DM density $\rho_{\rm DM}$, average density $\overline{\rho}_{\rm DM}$ and enclosed mass radial profiles for a system of virial mass $M_{\rm vir}=10^{14}$M$_{\rm \odot}$ at redshift $z=0$, with concentration given by D08, for three cases of the B10 profile. By definition, the virial radius $r_{\rm vir}$ is the same, regardless of the choice of $\beta$. If we assume that the concentration $c_{\rm vir}$ does not change for any $1 < \beta \leq 3$, the scale radius $r_{\rm s}$ is the same for all profiles. What changes instead is the characteristic density $\delta_{\rm c}$, as described in Section \ref{sec:Polytropic_gas_in_galaxy_groups_and_clusters}.

The DM average density and enclosed mass are higher (lower) than the NFW case in the $\beta>2$ ($\beta<2$) case for $r<r_{\rm vir}$, and lower (higher) than the NFW case in the $\beta>2$ ($\beta<2$) case for $r>r_{\rm vir}$. This results in different overdensity masses and radii: for $\beta>2$ ($\beta<2$), $M_{\rm \Delta_{\rm c}}$ and $r_{\rm \Delta_{\rm c}}$ are greater (smaller) than in the NFW case, for any $\Delta_{\rm c}>\Delta_{\rm vir}$ (which is the case at any redshift for $\Delta_{\rm c} \geq 200$).

For a system of known $M_{\rm vir}$ and known concentration, assuming that the baryonic mass is negligible, we can calculate the overdensity mass $M_{\rm \Delta_{\rm c}}$ and the overdensity radius $r_{\rm \Delta_{\rm c}}$, for any given value of $\Delta_{\rm c}$ and $\beta$. By combining the definitions of virial mass and overdensity mass with those of enclosed mass, we obtain $r_{\rm \Delta_{\rm c}}$ by numerically solving

\scalefont{0.94}
\begin{equation}
\frac{r_{\rm \Delta_{\rm c}}^3\Delta_{\rm c}}{r_{\rm vir}^3\Delta_{\rm vir}} = \frac{1+\left[(1-\beta)(r_{\rm \Delta_{\rm c}}/r_{\rm s})-1\right]\left(1+r_{\rm \Delta_{\rm c}}/r_{\rm s}\right)^{1-\beta}}{1+\left[(1-\beta)c_{\rm vir}-1\right]\left(1+c_{\rm vir}\right)^{1-\beta}}
\end{equation}
\normalsize

\noindent in the $\beta \neq 2$ case and

\begin{equation}
\frac{r_{\rm \Delta_{\rm c}}^3\Delta_{\rm c}}{r_{\rm vir}^3\Delta_{\rm vir}} = \frac{\ln{\left(1+r_{\rm \Delta_{\rm c}}/r_{\rm s}\right)}-(r_{\rm \Delta_{\rm c}}/r_{\rm s})/(1+r_{\rm \Delta_{\rm c}/r_{\rm s}})}{\ln{(1+c_{\rm vir})}-c_{\rm vir}/(1+c_{\rm vir})}
\end{equation}

\noindent in the $\beta=2$ case. We subsequently solve

\begin{equation}
\frac{M_{\rm \Delta_{\rm c}}}{M_{\rm vir}} = \frac{r_{\rm \Delta_{\rm c}}^3\Delta_{\rm c}}{r_{\rm vir}^3\Delta_{\rm vir}}
\end{equation}

\noindent to obtain $M_{\rm \Delta_{\rm c}}$.

In Figure \ref{fig:Overdensities}, we show the overdensity masses and radii as a function of virial mass, for four typical values of $\Delta_{\rm c}$ (including $\Delta_{\rm vir}$) and three values of $\beta$, when using the mass-concentration relation by D08 at redshift $z=0$ (when $\Delta_{\rm vir} \simeq 95$).

Given the weak dependence of concentration with mass (D08), we can fit simple\footnote{If we use mass-concentration relations with a stronger mass dependence (e.g. O10 and O11), the fitting relations are not this simple.} relations between $M_{\rm \Delta_{\rm c}}$, $r_{\rm \Delta_{\rm c}}$ and $M_{\rm vir}$:

\begin{equation}\label{eq:Mass_overdensity_fitting}
\frac{M_{\rm \Delta_{\rm c}}}{M_{\rm vir}}=M_{\rm 1}\log_{\rm 10}(M_{\rm vir})+M_{\rm 2}
\end{equation}

\noindent and

\begin{equation}\label{eq:Radius_overdensity_fitting}
\log_{\rm 10}(r_{\rm \Delta_{\rm c}})=R_{\rm 1}\log_{\rm 10}(M_{\rm vir})+R_{\rm 2},
\end{equation}

\noindent where the parameters $M_{\rm 1}$, $M_{\rm 2}$, $R_{\rm 1}$ and $R_{\rm 2}$ depend on $\Delta_{\rm c}$ and $\beta$, as shown in Table \ref{tab:Overdensities} for the $z=0$ case, when $r_{\rm \Delta_{\rm c}}$ is given in Mpc and $M_{\rm vir}$ is given in units of $10^{14}$M$_{\rm \odot}$. $R_{\rm 1-vir}=1/3$ simply from the definition of virial radius.

\renewcommand{\arraystretch}{1.3}
\begin{table}
\centering
\begin{tabular}{l l l l l}
\rowcolor[gray]{.85} $\beta=3/2$ &  $\Delta_{\rm c}=\Delta_{\rm vir}$ & $\Delta_{\rm c}=200$ & $\Delta_{\rm c}=500$ & $\Delta_{\rm c}=2500$ \\ [0.5ex]
$M_{\rm 1}$                      &  0                                 & -0.018259            & -0.029296            & -0.024567             \\
$M_{\rm 2}$                      &  1                                 & 0.701823             & 0.422358             & 0.130490              \\
$R_{\rm 1}$                      &  0.333333                          & 0.329559             & 0.323228             & 0.305437              \\
$R_{\rm 2}$                      &  0.080661                          & -0.077919            & -0.284488            & -0.690595             \\
\rowcolor[gray]{.85} NFW         &  $\Delta_{\rm c}=\Delta_{\rm vir}$ & $\Delta_{\rm c}=200$ & $\Delta_{\rm c}=500$ & $\Delta_{\rm c}=2500$ \\
$M_{\rm 1}$                      &  0                                 & -0.018785            & -0.035654            & -0.043279             \\
$M_{\rm 2}$                      &  1                                 & 0.800162             & 0.574959             & 0.257952              \\
$R_{\rm 1}$                      &  0.333333                          & 0.329928             & 0.324302             & 0.308514              \\
$R_{\rm 2}$                      &  0.080661                          & -0.058924            & -0.239741            & -0.591205             \\
\rowcolor[gray]{.85} $\beta=5/2$ &  $\Delta_{\rm c}=\Delta_{\rm vir}$ & $\Delta_{\rm c}=200$ & $\Delta_{\rm c}=500$ & $\Delta_{\rm c}=2500$ \\
$M_{\rm 1}$                      &  0                                 & -0.017403            & -0.037090            & -0.058379             \\
$M_{\rm 2}$                      &  1                                 & 0.871441             & 0.704620             & 0.408453              \\
$R_{\rm 1}$                      &  0.333333                          & 0.330437             & 0.325672             & 0.312246              \\
$R_{\rm 2}$                      &  0.080661                          & -0.046555            & -0.210196            & -0.523894             \\ [1ex]
\end{tabular}
\caption{Overdensity fitting parameters. We list the fitting parameters for Equations \eqref{eq:Mass_overdensity_fitting} and \eqref{eq:Radius_overdensity_fitting}, for systems at redshift $z=0$, with concentration given by D08, for four typical values of $\Delta_{\rm c}$ -- 200, 500, 2500 and $\Delta_{\rm vir}(z=0)\simeq 95$ -- and for three values of $\beta$ -- 3/2, 2 (NFW) and 5/2 -- in the B10 profile, when $r_{\rm \Delta_{\rm c}}$ is given in Mpc and $M_{\rm vir}$ is given in units of $10^{14}$M$_{\rm \odot}$.}
\label{tab:Overdensities}
\end{table} 
\renewcommand{\arraystretch}{1.0}

%\newpage
%\hspace{1cm}
%\newpage

\section{The cooling function}\label{sec:Cooling_Function}

\begin{figure}
\centering
\vspace{5pt}
\includegraphics[width=0.95\columnwidth,angle=0]{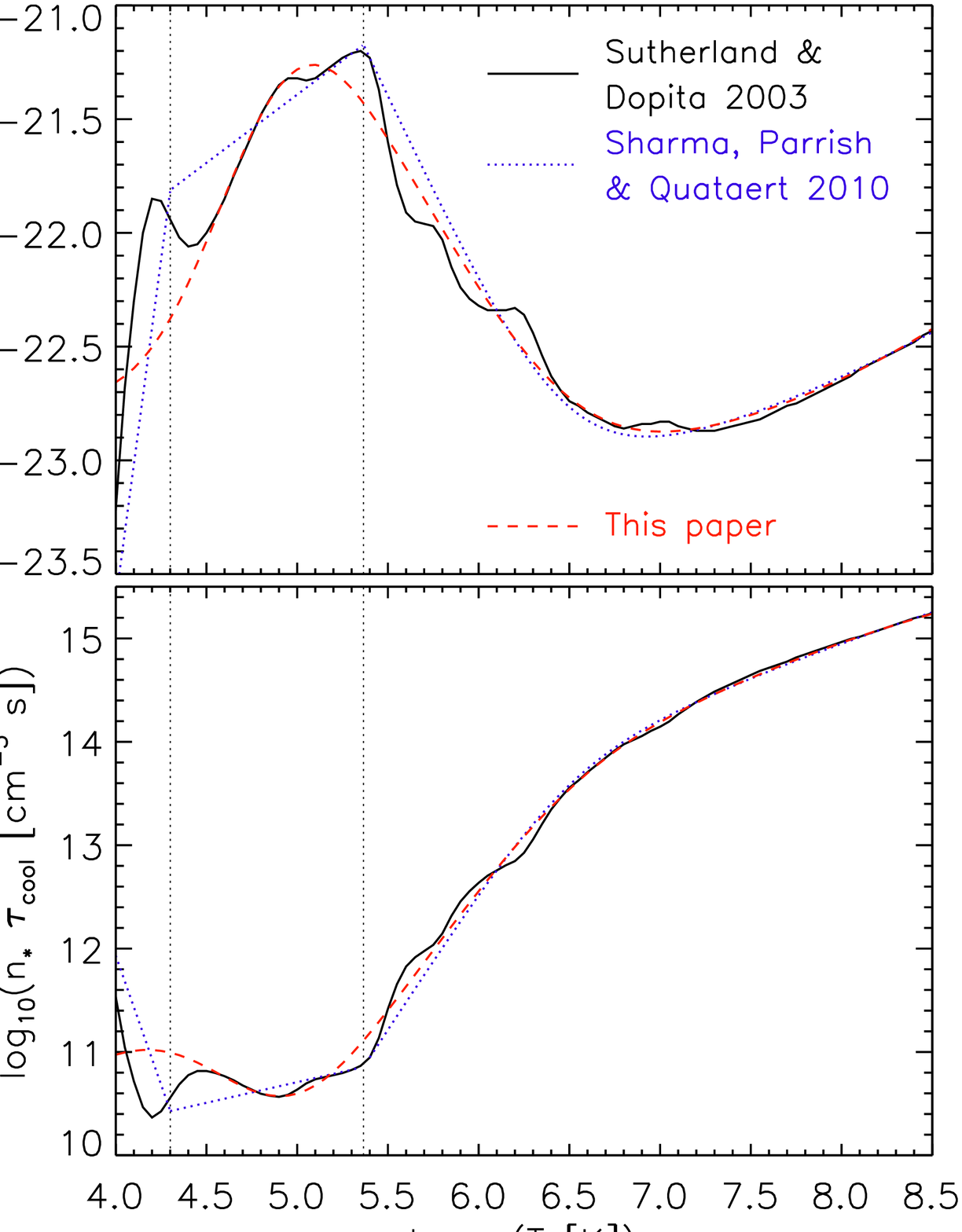}
\vspace{13pt}
\caption{Cooling function (top panel) and normalised cooling time (bottom panel) for an optically-thin, fully-ionized, low-density gas of mean metallicity [Fe/H] $=-0.5$ in collisional ionization equilibrium. \emph{Black, solid lines}: data from Sutherland \& Dopita (1993). \emph{Blue, dotted lines}: fitting function to the data by Sharma et al. (2010). \emph{Red, dashed lines}: our fitting function to the data. The two \emph{black, vertical, dotted lines} denote the temperature regimes of Equation \eqref{eq:Cooling_SPQ2010}.}
\label{fig:Cooling}
\end{figure}

Cooling, defined as the total energy lost per unit volume per unit time, can be usually parametrized in the optically-thin, fully-ionized, low-density limit as $\Lambda_{\rm net}=n_{\rm i}n_{\rm e}\Lambda_{\rm N}$, where the cooling function $\Lambda_{\rm N}$ is defined below. We assume, following Tozzi \& Norman (2001: TN01), $n_{\rm i}n_{\rm e}=0.704\left(\rho/m_{\rm p}\right)^2$ for a gas with mean metallicity [Fe/H] $=-0.5$.

Cooling time, defined as the total energy of the gas divided by the energy loss, is then

\begin{equation}
\tau_{\rm cool}=\frac{(n_{\rm e}+n_{\rm i})k_{\rm B}T}{(\gamma-1)n_{\rm e}n_{\rm i}\Lambda_{\rm N}}=\frac{k_{\rm B}T}{n_*(\gamma-1)\Lambda_{\rm N}},
\end{equation}

\noindent where $n_* \equiv n_{\rm e}n_{\rm i}/(n_{\rm e}+n_{\rm i})$ and we have assumed that electron and ion temperatures are equal to each other.

Under the assumption of collisional ionization equilibrium, we adopt the cooling function by Sutherland \& Dopita (1993). TN01 first provided the fitting function, valid only for $k_{\rm B}T>0.02$ keV,

\begin{equation}\label{eq:Cooling_TN2001}
\Lambda_{\rm N} = C_{\rm 1}(k_{\rm B}T)^{\alpha}+C_{\rm 2}(k_{\rm B}T)^{\beta}+C_{\rm 3},
\end{equation}

\noindent where the constants $C_{\rm 1}$, $C_{\rm 2}$ and $C_{\rm 3}$ depend on the mean metallicity of the gas, $\alpha=-1.7$ and $\beta=0.5$. For a gas with mean metallicity [Fe/H] $=-0.5$, the constants are $C_{\rm 1}=8.6 \times 10^{-25}$ erg cm$^3$ s$^{-1}$ keV$^{-\alpha}$, $C_{\rm 2}=5.8 \times 10^{-24}$ erg cm$^3$ s$^{-1}$ keV$^{-\beta}$, and $C_{\rm 3}=6.3 \times 10^{-24}$ erg cm$^3$ s$^{-1}$.

Sharma, Parrish, \& Quataert (2010) recently generalised the fitting function of TN01 in the [Fe/H] $=-0.5$ case, by considering three separate temperature regimes:

\begin{equation}\label{eq:Cooling_SPQ2010}
\Lambda_{\rm N} =
         \begin{cases}\displaystyle{A_{\rm 1}(k_{\rm B}T)^{A_{\rm 2}}} \hspace{3.05cm} \mbox{if $T < T_1$,} \cr
                      \displaystyle{B_{\rm 1}(k_{\rm B}T)^{B_{\rm 2}}} \hspace{2.28cm} \mbox{if $T_2 \geq T \geq T_1$,} \cr
		      \displaystyle{C_{\rm 1}(k_{\rm B}T)^{\alpha}+C_{\rm 2}(k_{\rm B}T)^{\beta}+C_{\rm 3}} \hspace{0.62cm} \mbox{if $T > T_2$,} \cr
         \end{cases}
\end{equation}

\noindent where $k_{\rm B}T_1=0.0017235$ keV, $k_{\rm B}T_2=0.02$ keV, the constants $C_{\rm 1}$, $C_{\rm 2}$, $C_{\rm 3}$, $\alpha$ and $\beta$ are the same as in TN01 and the other constants are $A_{\rm 1}=5.891 \times 10^{-6}$ erg cm$^3$ s$^{-1}$ keV$^{-A_{\rm 2}}$, $B_{\rm 1}=7.027 \times 10^{-21}$ erg cm$^3$ s$^{-1}$ keV$^{-B_{\rm 2}}$, $A_{\rm 2}=6$ and $B_{\rm 2}=0.6$.

We adopt instead a new fitting function, which has the advantage of being a single formula like the one by TN01, for easier computational and/or analytical use, and at the same time has the advantage of taking into account the decrease of $\Lambda_{\rm N}$ at low temperatures (for $\log_{\rm 10}(T\hbox{[K]})>4.45$) and the relativistic correction for the Bremsstrahlung emission:

\begin{eqnarray}\label{eq:Cooling_this_paper}
\Lambda_{\rm N} &=& C_{\rm 1}(k_{\rm B}T)^{\alpha}\exp\left[-\left(\frac{C_{\rm 4}}{1+C_{\rm 5}k_{\rm B}T}\right)^{C_{\rm 6}}\right]+\\
&+& C_{\rm 2}(k_{\rm B}T)^{\beta}(1+C_{\rm 7}k_{\rm B}T)+C_{\rm 3}. \nonumber
\end{eqnarray}

In Figure \ref{fig:Cooling}, we show the cooling function and normalised cooling time, $n_* \tau_{\rm cool}$, as a function of temperature, for the [Fe/H] $=-0.5$ case. For comparison, we show the original data from Sutherland \& Dopita (1993), the fitting function by Sharma et al. (2010) and the new fitting function introduced in this paper.

In Table \ref{tab:Cooling_Function}, we list the fitting parameters for Equation \eqref{eq:Cooling_this_paper}, for a mean metallicity of [Fe/H] $=-0.5$, for both the TN01 and this paper cases.

\renewcommand{\arraystretch}{1.3}
\begin{table}
\centering
\begin{tabular}{l l l l}
\rowcolor[gray]{.85}      & TN01                      & This paper                    & Units                                 \\ [0.5ex]
$C_{\rm 1}$               & $8.6\times10^{-25}$       & $1.441\times10^{-24}$         & [erg cm$^3$ s$^{-1}$ keV$^{-\alpha}$] \\
$C_{\rm 2}$               & $5.8\times10^{-24}$       & $5.247\times10^{-24}$         & [erg cm$^3$ s$^{-1}$ keV$^{-\beta}$]  \\
$C_{\rm 3}$               & $6.3\times10^{-24}$       & $6.702\times10^{-24}$         & [erg cm$^3$ s$^{-1}$]                 \\
$\alpha$                  & -1.7                      & -1.443                        & [dimensionless]                       \\
$\beta$                   & $5.0\times10^{-1}$        & $5.0\times10^{-1}$            & [dimensionless]                       \\
$C_{\rm 4}$               & 0                         & 1.449                         & [dimensionless]                       \\
$C_{\rm 5}$               & 0                         & $5.268 \times 10^1$           & [keV$^{-1}$]                          \\
$C_{\rm 6}$               & 0                         & 6.303                         & [dimensionless]                       \\
$C_{\rm 7}$               & 0                         & $5.0 \times 10^{-3}$          & [keV$^{-1}$]                          \\
$C_{\rm 1}'$              & $8.8\times10^{-13}$       & $2.256\times10^{-14}$         & [erg cm$^3$ s$^{-1}$ K$^{-\alpha'}$]  \\
$C_{\rm 2}'$              & $1.7\times10^{-27}$       & $1.540\times10^{-27}$         & [erg cm$^3$ s$^{-1}$ K$^{-\beta'}$]   \\
$C_{\rm 3}'$              & $6.3\times10^{-24}$       & $6.702\times10^{-24}$         & [erg cm$^3$ s$^{-1}$]                 \\
$\alpha'$                 & -1.7                      & -1.443                        & [dimensionless]                       \\
$\beta'$                  & $5.0\times10^{-1}$        & $5.0\times10^{-1}$            & [dimensionless]                       \\
$C_{\rm 4}'$              & 0                         & 1.449                         & [dimensionless]                       \\
$C_{\rm 5}'$              & 0                         & $4.551\times10^{-6}$          & [K$^{-1}$]                            \\
$C_{\rm 6}'$              & 0                         & 6.303                         & [dimensionless]                       \\
$C_{\rm 7}'$              & 0                         & $4.309\times10^{-10}$         & [K$^{-1}$]                            \\ [1ex]
\end{tabular}
\caption{Cooling function fitting parameters. We list the fitting parameters for Equation (B3), for a mean metallicity of [Fe/H] $=-0.5$, for both the TN01 and this paper cases. When substituting $k_{\rm B}T$ (in keV) with $T$ (in K), the primed parameters in the table should be used instead.}
\label{tab:Cooling_Function}
\end{table} 
\renewcommand{\arraystretch}{1.0}

\end{document}